\documentclass{article}

\PassOptionsToPackage{numbers}{natbib}

\usepackage[preprint]{neurips_2024}

\usepackage[utf8]{inputenc} 
\usepackage[T1]{fontenc}    
\usepackage{hyperref}       
\usepackage{url}            
\usepackage{booktabs}       
\usepackage{amsfonts}       
\usepackage{nicefrac}       
\usepackage{microtype}      
\usepackage{xcolor}         
\usepackage{amsmath}
\usepackage{graphicx}
\usepackage{multirow}
\usepackage{enumitem}
\usepackage{amsthm}
\usepackage{mathtools}
\usepackage{subfigure}
\usepackage{footmisc}
\usepackage{wrapfig}
\usepackage{bbm}
\usepackage{float}
\usepackage{colortbl}
\usepackage{nicematrix}
\usepackage{tabularx}
\usepackage{caption}
\usepackage{array}
\usepackage{subcaption}

\hypersetup{
    colorlinks=true,
    citecolor=blue,
    linkcolor=blue,
    filecolor=magenta,      
    urlcolor=blue,
}

\usepackage{xspace}
\usepackage{lipsum}
\usepackage{tikz}
\usetikzlibrary{tikzmark}

\usepackage{threeparttable}
\usepackage{rotating}

\usepackage{wrapfig} 

\newcolumntype{b}{X}
\newcolumntype{s}{>{\hsize=.2\hsize}X}

\newlength\savewidth
\newcommand\shline{\noalign{\global\savewidth\arrayrulewidth
                            \global\arrayrulewidth 1.5pt}%
                   \hline
                   \noalign{\global\arrayrulewidth\savewidth}
                   }

\title{Generative Pre-Trained Diffusion Paradigm for Zero-Shot Time Series Forecasting}

\author{
	Jiarui Yang\textsuperscript{\rm 1, \rm 2}, 
	Tao Dai\textsuperscript{\rm 3}, 
	Naiqi Li\textsuperscript{\rm 2}, 
	Junxi Wu\textsuperscript{\rm 1,\rm 2}, 
    Peiyuan Liu\textsuperscript{\rm 2},
    Jigang Bao\textsuperscript{\rm 2},\\
    \textbf{Jinmin Li\textsuperscript{\rm 2}},
     \textbf{Haigang Zhang\textsuperscript{\rm 4}},
     \textbf{Shu-Tao Xia\textsuperscript{\rm 2}},\\
     \\
    \textsuperscript{\rm 1}Nankai University.\\
	\textsuperscript{\rm 2}Tsinghua Shenzhen International Graduate School, Tsinghua University.\\
    \textsuperscript{\rm 3}Shenzhen University. \textsuperscript{\rm 4}Shenzhen Polytechnic University.}

\begin{document}

\maketitle

\begin{abstract}

In recent years, generative pre-trained paradigms such as Large Language Models (LLMs) and Large Vision Models (LVMs) have achieved revolutionary advancements and widespread real-world applications. Particularly, the emergence of pre-trained LLMs-based temporal works, compared to previous deep model approaches, has demonstrated superior generalization and robustness, showcasing the potential of generative pre-trained paradigms as foundation models for time series. However, those LLMs-based works mainly focus on cross-modal research, i.e., leveraging the language capabilities of LLMs in time series contexts. Although they have achieved impressive performance, there still exist the issues of concept drift caused by differences in data distribution and inflexibility caused by misalignment of dimensions. To this end, inspired by recent work on LVMs, we reconsider the paradigm of time series modeling. In this paper, we comprehensively explore, for the first time, the effectiveness and superiority of the Generative Pre-trained Diffusion (GPD) paradigm in real-world multivariate time series forecasting (TSF). Specifically, to mitigate performance bias introduced by sophisticated networks, we propose a straightforward MLP diffusion network for unconditional modeling of time series. Then we employ a zero-shot and tuning-free method to predict (generate) future data using historical data as prompts. The GPD paradigm is established on the time series modality, effectively preventing the phenomenon of concept drift, and enabling flexible forecasting of arbitrary lengths. We demonstrate that the GPD paradigm achieves comprehensive performance and generalization comparable to current SOTA LLM-based and deep model paradigms on mainstream benchmarks and various TSF tasks. Extensive experiments validate the potential of the GPD paradigm and its assistance in future related research.

\end{abstract}

\section{Introduction}
\label{Intro}

In domains crucial to our daily lives such as finance, healthcare, and transportation, vast amounts of irregular time series data abound, characterized by varying lengths, features, and distributions~\citep{ yaffee2000introduction,pincus2004irregularity}. Time series forecasting (TSF) have long held significant importance and garnered considerable attention in these real-world domains~\citep{lim2021time}. In a nutshell, TSF is described as given a historical sequence $Y_H=\{y^{0},y^{1}, ..., y^{H} | y^i \in R^D \}$ of length $H$ with $D$ features, predicting future sequence $Y_P=\{y^{H+1},y^{H+2}, ..., y^{L} | y^i \in R^D\}$ of length $P$, where $L=P+H$, represents the total length. Confronted with the complexity of real-world time series, TSF algorithms usually require heightened generalization, flexibility, and robustness~\citep{kaastra1996designing,stefenon2023aggregating}.

\begin{figure*}[t]
    \centering
    \includegraphics[height=0.30\textwidth]{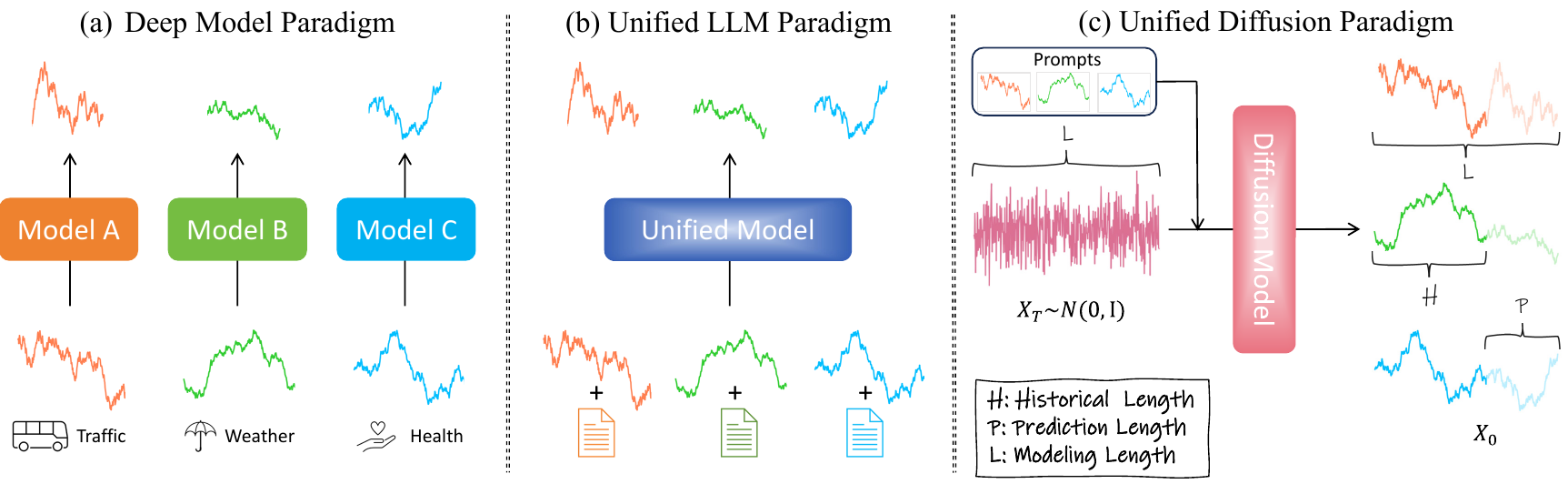}

    \caption{(a) The exclusive model establishes a singular mapping from history to future within a specific domain. (b) The unified model leverages the capabilities of LLMs and domain-specific textual instructions to differentiate and construct complex mappings. (c) The diffusion model establishes global underlying statistical characteristics across different domains. Note that paradigms (a) and (b) require retraining when altering the length of historical context. The diffusion paradigm offers flexibility in adjusting the length of historical prompts.}
    \label{para}
    \vspace{-0.7cm}
\end{figure*}

Over the past decade, with the continuous development of deep networks, deep models~\citep{hewamalage2021recurrent,chen2023long,qin2024mambavc} have achieved significant success in TSF. Particularly, the Transformer-based deep models paradigm have recently come to dominate the field~\citep{patchtst,informer,autoformer}. However, the emergence of large models and large-scale data has revealed the limitations of the traditional paradigm~\citep{unitime}. Firstly, as depicted in Fig. \ref{para}(a), the deep models paradigm is typically trained and tested on small-scale networks and single-domain small data. This mechanism may be overly restrictive for real-world applications, resulting in compromised generalization and robustness of models. Secondly, these models usually have fixed input-output dimensions, requiring fixed-length look-back windows during both training and testing. Consequently, they struggle to handle real-world time series of irregular lengths flexibly.

Foundation models of Natural Language Processing (NLP)~\citep{llama,gpt3} and Computer Vision (CV)~\citep{sora,bai2023sequential,sd} are rapidly gaining prominence, demonstrating remarkable generalization and robustness across various real-world scenarios. The inherent sequence modeling capability of LLMs naturally lends itself to time series modeling characteristics. Consequently, recent efforts~\citep{unitime,llmtime,timellm,gpt4ts} in LLM-based TSF have achieved preliminary success. However, much of the LLMs-based works still follows the previous deep models paradigm. Smaller datasets and single-domain learning do not effectively showcase the superiority of LLMs in the context of time series. To address this, \citet{unitime} propose UniTime, which empowers LLMs with cross-domain data, as shown in Fig. \ref{para}(b). UniTime is the first to validate the more robust generalization and performance brought about by cross-domain learning. The success of LLMs stems from adequately pre-training and learning universal language knowledge representations. However, recent LLMs-based works have failed to effectively address the issue of pre-trained weight distortion~\citep{shen2023cross, llata}, inevitably introducing the phenomenon of concept drift and consequently leading to sub-optimal performance. Moreover, due to modeling constraints, most LLMs-based methods remain inflexible, requiring fixed input and output lengths similar to deep models paradigm. Yet, a sound foundation model should be as flexible as ChatGPT in engaging in dynamic conversations, accommodating input of arbitrary lengths.

\begin{wrapfigure}{r}{0.38\textwidth}
  \centering
  \vspace{-10pt}
  \includegraphics[width=0.38\textwidth]{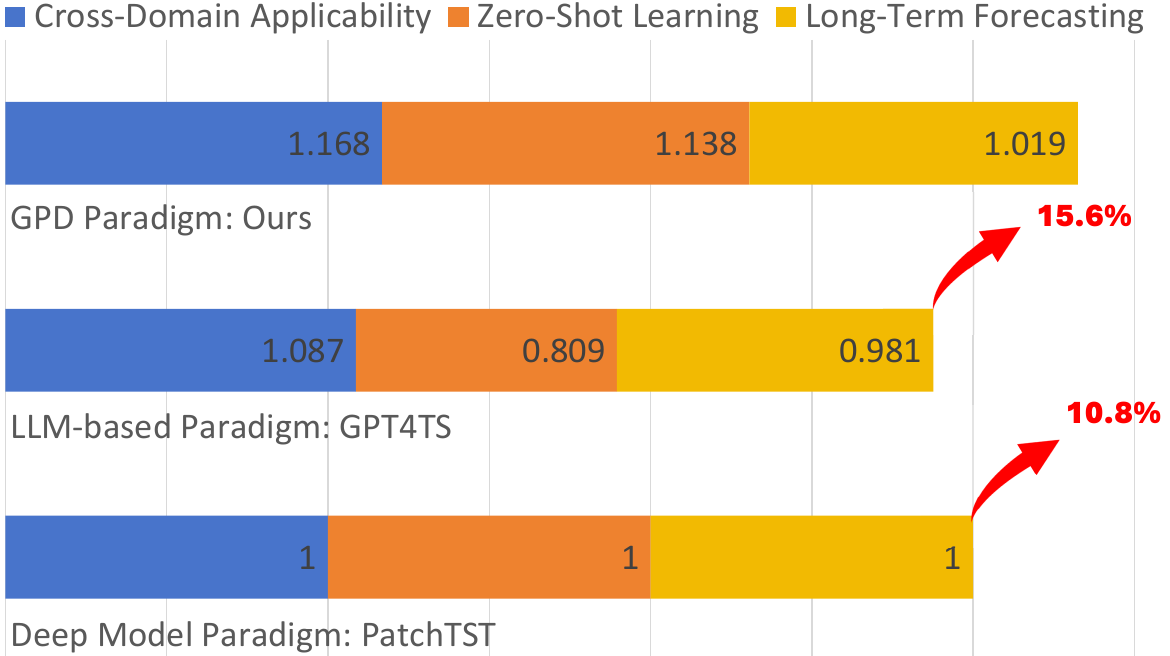}
  \vspace{-10pt}
  \caption{Comparison of comprehensive performance of different paradigms.}
  \label{fig:comprehensive}
\end{wrapfigure}

Recently, the video generation paradigm exemplified by Sora~\citep{sora,dit} validates the potential of diffusion models~\citep{sd,ddpm} in modeling multivariate time series. The application of pre-trained video generators to downstream tasks, such as video frame prediction~\citep{davtyan2023efficient,blattmann2023stable}, is particularly inspiring to us. To this end, we rethink the modeling paradigm and explore the possibility and superiority of Generative Pre-trained Diffusion (GPD) paradigm in TSF. Firstly, as shown in Fig. \ref{fig:comprehensive}, the GPD paradigm demonstrates comprehensive performance comparable to LLM-based and deep model paradigms. Secondly, unlike the cross-modal approach of LLMs paradigm, we construct the foundation model based on the time series modality. Therefore, our method have good interpretability, thereby avoiding conceptual drift. Finally, as shown in Fig. \ref{para}(c), the GPD paradigm can flexibly apply any historical and future lengths ($H\&P$) within the maximum length $L$.

While there exist some temporal works based on diffusion models~\citep{timegrad,csdi,sssd,d3vae,ldt}, most of them rely on additional predictors such as initial estimates provided by RNNs. This modeling approach imposes constraints on the model, compromising prior knowledge and robustness~\citep{wang2023exploiting}. The joint training of additional predictor modules with diffusion models leads to a potential coupling between priors and predictors. This coupling results in a trade-off between diffusion denoising and predictor performance, making it challenging to effectively apply them to downstream tasks. Unlike multi-modal text or image embeddings where inputs have close semantic associations~\citep{clip}, the initial estimates from predictors inherently contain errors compared to ground truth, which are gradually amplified during the sampling stage. Diverging from the majority of diffusion-based temporal works, we adopt a pre-trained perspective, constructing an unconditional multivariate time series generator as the foundation model. In downstream TSF tasks, we utilize any length $H$ of historical sequence as a prompt to generate future sequence of length $P$ in a zero-shot manner at once. The superiority of the GPD paradigm lies in its avoidance of learning the underlying mapping between input and output, instead effectively leveraging the underlying statistical features and latent data distribution of time series data. Consequently, GPD exhibits advantages in terms of generalization and universality. We summarize the main contributions of this paper as follows:

\begin{itemize}

    \item We propose a novel and exploratory Generative Pre-trained Diffusion (GPD) paradigm for time series analysis. Our experiments and methodology offer valuable guidance and inspiration for future related research endeavors.

    \item Diffusion-based method beats methods based on LLMs and deep models in multiple forecasting tasks for the first time.

    \item Our proposed GPD supports forecasting with arbitrary lengths of historical data, demonstrating higher flexibility and generality compared to prior methods.

\end{itemize}

\section{Method}
\subsection{Preliminary}
Denoising Diffusion Probabilistic Models (DDPMs~\citep{ddpm,ddim}) are a class of probability-based generative models that learn the mapping from high-dimensional Gaussian noise to the data distribution, enabling the generation of novel samples. From a mathematical perspective, DDPMs can be formalized as a Markov chain process~\citep{puterman1990markov}. Specifically, the process starts from the data distribution $X \sim q(X)$, and gradually adds Gaussian noise according to a pre-defined schedule $\{\beta_0, ..., \beta_T\}$, generating a series of intermediate states $\{X_0, ..., X_T\}$ defined by the conditional probability distributions $q(X_t|X_{t-1})$. Here, $X_T$ approximates standard Gaussian noise. This process is called the forward diffusion, and it is Gaussian:
\begin{equation}
    q(X_{t}|X_{t-1}) = \mathcal{N}(X_{t}; \sqrt{\alpha_{t}}X_{t-1}, (1-\alpha_{t})\mathbf{I})
    \label{eq1}
\end{equation}
where $\alpha_t = 1 - \beta_t$. The generation process then runs in reverse order, reconstructing the data from noise, by sampling according to the inverse conditional probabilities $p_\theta(X_{t-1}|X_t)$. This process is called the reverse diffusion, and it is defined as:
\begin{equation}
    p_\theta(X_{t-1}|X_{t}) = \mathcal{N}(X_{t-1}; \mu_{\theta}(X_{t}, t), \tilde \beta_t \mathbf{I})
    \label{eq2}
\end{equation}

where $\tilde \beta_t$ is a fixed variance, and $\mu_\theta$ is the mean, which contains the learnable parameter $\theta$. Essentially, we optimize the parameter $\theta$ by minimizing the Evidence Lower Bound (ELBO~\citep{sohl2015deep}) of the estimated negative log-likelihood $\mathbb{E} \left[- \text{log} p_{\theta}(X_0)\right]$, in order to maximize the probability of observing the training sample $X_0$ estimated by $p_{\theta}(X_0)$.

\subsection{Pre-Trained Diffusion Model}
We focus on exploring the potential of Generative Pre-trained Diffusion (GPD) paradigm in zero-shot time series forecasting (TSF). Therefore, to avoid performance bias introduced by sophisticated networks, we opt for a simple MLP network as a baseline for modeling. We adopt the MLP network proposed by~\citet{diffae} which consists of multiple MLP blocks concatenated together, with each block having a skip connection from the input. We adhere to the channel-wise independent setting of UniTime~\citep{unitime} and PatchTST~\citep{patchtst} to flexibly handle data from different domains with different feature. Relative to optimizing the ELBO, we opt for the simplified objective function~\citep{ddpm}:
\begin{equation}
    \mathcal{L}_{\text{simple}} = \mathbb{E}_{\Gamma, t}[||\Gamma_{\theta}(X_t, t)-\Gamma||^2_2]
    \label{eq3}
\end{equation}

where $\Gamma$ corresponds to the noise $\epsilon$ or initial state $X_0$ predicted by the network. 

\subsection{Zero-Shot Prompt Forecasting}
\label{sec:prompt foecasting}

Regression models typically learn a deterministic and unique mapping relationship from historical data. In contrast, diffusion models can take a more global perspective $\{y^i\}_{i=1}^L$ and explore multiple approximate solutions in a broader prior space. We utilize prompts similar to those used in the image generation domain, such as video frame prediction~\citep{blattmann2023stable} and image inpainting~\citep{song2020score}, to predict the posterior distribution at each sampling step by injecting the historical data $\{y^i\}_{i=1}^H$ as the prompt. This requires aligning the prompt distribution with the pre-trained model's posterior distribution through posterior sampling~\citep{ilvr}. We first apply forward diffusion to the historical prompt to add noise:

\begin{equation}
    Y_{H_t} = \sqrt{\bar \alpha_t} Y_{H_0} + \sqrt{1 - \bar \alpha_t} {\epsilon}_{\theta}(X_t,t)
    \label{eq4}
\end{equation}
where $Y_{H_0}=\{y_0^i\}_{i=1}^H$ represents the GT historical sequence of length $H$, and $\epsilon_\theta$ denotes the noise predicted by the pre-trained model. Then, we align the distributions through Eq. \ref{eq2}:
\begin{equation}
    Y_{H_{t-1}}=p_\theta(Y_{H_{t-1}}|Y_{H_{t}}) = \mathcal{N}(Y_{H_{t-1}}; \mu_{\theta}(Y_{H_{t}}, t), \tilde \beta_t \mathbf{I})
    \label{eq5}
\end{equation}

At each time step of diffusion sampling, we replace the corresponding sequence of the pre-trained model's posterior distribution $X_{t-1}$ with $Y_{H_{t-1}}$:

\begin{equation}
    X_{H_{t-1}}=:\{x_{t-1}^i\}_{i=1}^H = Y_{H_{t-1}}
    \label{eq6}
\end{equation}

We ultimately sample a time series $\hat X_{0}$ of length $L$ constructed from historical data and future predictions. In fact, our approach relies entirely on a pre-trained generative model without fine-tuning for downstream forecasting tasks. Therefore, to alleviate randomness and enhance zero-shot TSF accuracy, we perform multiple samplings and average them to obtain an unbiased estimate:

\begin{equation}
    E[\hat X_0] = \sum_{i=1}^{n} \hat X_{0}^i \cdot p_\theta(\hat X_{0}^i|X_T^i)
\end{equation}

\begin{table}[htbp]
\vspace{-0.5cm}
\small
  \centering
  \caption{\textbf{Benchmarks for Different Tasks.}}
  \setlength{\tabcolsep}{2pt}
  \begin{tabular}{cccc}
    \toprule
    Tasks & Datasets & Input Length & Output Length \\
    \midrule
    Cross-Domain & ETT (4 subsets), Electricity, Weather, Exchange~\citep{lai2018modeling} & 96 & 96$\sim$720 \\
    Long-term Forecasting & ETT (4 subsets), Electricity~\citep{misc_electricityloaddiagrams20112014_321}, Weather, Traffic~\citep{misc_pems-sf_204} & 512 & 96$\sim$720 \\
    Short-term Forecasting & Electricity, Solar~\citep{lai2018modeling}, Traffic, Taxi~\citep{timegrad}, Wikipedia~\citep{gluonts_arxiv}& 48$\sim$168 & 24$\sim$30 \\
    Zero-Shot & ETT (4 subsets)~\citep{informer} & 512 & 96$\sim$720 \\
    \bottomrule
  \end{tabular}
  \label{tab:benchmarks}
\end{table}

\section{Experiments}
\subsection{Datasets and Baselines}

We extensively evaluate the GPD paradigm on 9 real-world benchmarks, covering cross-domain, long-term and short-term, and zero-shot forecasting. The specific dataset settings and sequence lengths are shown in Table \ref{tab:benchmarks}. The comprehensive baselines evaluated include:

LLM-based paradigms: \textit{UniTime (WWW'24)}~\citep{unitime}, \textit{TimeLLM (ICLR'24)}~\citep{timellm}, \textit{LLMTime (NeurIPS'23)}~\citep{llmtime}, \textit{GPT4TS (NeurIPS'23)}~\citep{gpt4ts} and \textit{LLM4TS (ArXiv'23)}~\citep{llm4ts}.

Deep model paradigms: \textit{PatchTST (ICLR'23)}~\citep{patchtst}, \textit{TimesNet (ICLR'23)}~\citep{timesnet}, \textit{DLinear (AAAI'23)}~\citep{dlinear}, \textit{FEDformer (ICML'22)}, \textit{Autoformer (NeurIPS'21)}~\citep{autoformer}, \textit{Informer (AAAI'21)}~\citep{informer} and \textit{Pyraformer (ICLR'21)}~\citep{pyraformer}.

Diffusion paradigms: \textit{LDT (AAAI'24)}~\citep{ldt}, \textit{$\text{D}^{3}\text{VAE}$~\citep{d3vae} (NeurIPS'22)}, \textit{SSSD (TML'22)}~\citep{sssd}, \textit{CSDI (NeurIPS'21)}~\citep{csdi} and \textit{TimeGrad (ICML'21)}~\citep{timegrad}.

\begin{table*}[h]
    \centering
    \small
    \tabcolsep=0.8mm
    \renewcommand\arraystretch{1.4}
    \caption{\textbf{Cross-domian forecasting performance comparisons}.  Avg is averaged over all predictive lengths. We highlight the best performance across datasets with boldface to the left of the double vertical lines for the cross-dataset trained models, and we boldface and underscore the best overall performance on the entire row.}
    \vspace{0.2em}
    \resizebox{\textwidth}{!}{
    \begin{tabular}{cc|cc|cc|cc|cc||cc|cc|cc|cc|cc|cc|cc}
    \shline
    \multicolumn{2}{c}{\multirow{3}{*}{Method}} & \multicolumn{8}{c||}{\textbf{\textit{Models Trained Across Datasets}}} & \multicolumn{14}{c}{\textbf{\textit{Models Trained on Each Dataset}}} \\ 
    
    \cline{3-24}
    \multicolumn{2}{c}{} & \multicolumn{2}{c|}{Ours} & \multicolumn{2}{c|}{UniTime} & \multicolumn{2}{c|}{GPT4TS$^{\dag}$} & \multicolumn{2}{c||}{PatchTST$^{\dag}$} & \multicolumn{2}{c|}{GPT4TS$^*$} & \multicolumn{2}{c|}{PatchTST$^*$} & \multicolumn{2}{c|}{TimesNet} & \multicolumn{2}{c|}{DLinear} & \multicolumn{2}{c|}{FEDformer} & \multicolumn{2}{c|}{Autoformer} & \multicolumn{2}{c}{Informer} \\ \cline{3-24} 
    \multicolumn{2}{c}{} & MSE & MAE & MSE & MAE & MSE & MAE & MSE & MAE & MSE & MAE & MSE & MAE & MSE & MAE & MSE & MAE & MSE & MAE & MSE & MAE & MSE & MAE \\
    \hline \hline
    \multicolumn{1}{c|}{\multirow{5}{*}{\rotatebox{90}{ETTm1}}} & 96 & 0.465 & 0.458 & \textbf{\underline{0.322}} & \textbf{\underline{0.363}} & 0.509 & 0.463 & 0.927 & 0.604 & 0.335 & 0.369 & 0.344 & 0.373 & 0.338 & 0.375 & 0.345 & 0.372 & 0.379 & 0.419 & 0.505 & 0.475 & 0.672 & 0.571 \\
    \multicolumn{1}{c|}{} & 192 & 0.493 & 0.473 & \textbf{\underline{0.366}} & \textbf{0.387} & 0.537 & 0.476 & 0.964 & 0.620 & 0.374 & \textbf{\underline{0.385}} & 0.367 & 0.386 & 0.374 & 0.387 & 0.380 & 0.389 & 0.426 & 0.441 & 0.553 & 0.496 & 0.795 & 0.669 \\
    \multicolumn{1}{c|}{} & 336 & 0.508 & 0.481 & \textbf{0.398} & \textbf{0.407} & 0.564 & 0.488 & 1.041 & 0.656 & 0.407 & \textbf{\underline{0.406}} & \textbf{\underline{0.392}} & 0.407 & 0.410 & 0.411 & 0.413 & 0.413 & 0.445 & 0.459 & 0.621 & 0.537 & 1.212 & 0.871 \\
    \multicolumn{1}{c|}{} & 720 & 0.552 & 0.504 & \textbf{\underline{0.454}} & \textbf{\underline{0.440}} & 0.592 & 0.504 & 0.950 & 0.636 & 0.469 & 0.442 & 0.464 & 0.442 & 0.478 & 0.450 & 0.474 & 0.453 & 0.543 & 0.490 & 0.671 & 0.561 & 1.166 & 0.823 \\ \cline{2-24} 
    \multicolumn{1}{c|}{} & Avg & 0.505 & 0.479 & \textbf{\underline{0.385}} & \textbf{\underline{0.399}} & 0.551 & 0.483 & 0.971 & 0.629 & 0.396 & 0.401 & 0.392 & 0.402 & 0.400 & 0.406 & 0.403 & 0.407 & 0.448 & 0.452 & 0.588 & 0.517 & 0.961 & 0.734 \\
    \hline \hline
    \multicolumn{1}{c|}{\multirow{5}{*}{\rotatebox{90}{ETTm2}}} & 96 & 0.221 & 0.295 & \textbf{0.183} & \textbf{0.266} & 0.229 & 0.304 & 0.240 & 0.318 & 0.190 & 0.275 & \textbf{\underline{0.177}} & \textbf{\underline{0.260}} & 0.187 & 0.267 & 0.193 & 0.292 & 0.203 & 0.287 & 0.255 & 0.339 & 0.365 & 0.453 \\
    \multicolumn{1}{c|}{} & 192 & 0.278 & 0.329 & \textbf{0.251} & \textbf{0.310} & 0.287 & 0.338 & 0.301 & 0.352 & 0.253 & 0.313 & \textbf{\underline{0.246}} & \textbf{\underline{0.305}} & 0.249 & 0.309 & 0.284 & 0.362 & 0.269 & 0.328 & 0.281 & 0.340 & 0.533 & 0.563 \\
    \multicolumn{1}{c|}{} & 336 & 0.332 & 0.359 & \textbf{0.319} & \textbf{0.351} & 0.337 & 0.367 & 0.367 & 0.391 & 0.321 & 0.360 & \textbf{\underline{0.305}} & \textbf{\underline{0.343}} & 0.321 & 0.351 & 0.369 & 0.427 & 0.325 & 0.366 & 0.339 & 0.372 & 1.363 & 0.887 \\
    \multicolumn{1}{c|}{} & 720 & 0.423 & \textbf{0.409} & \textbf{0.420} & 0.410 & 0.430 & 0.416 & 0.451 & 0.432 & 0.411 & 0.406 & 0.410 & 0.405 & \textbf{\underline{0.408}} & \textbf{\underline{0.403}} & 0.554 & 0.522 & 0.421 & 0.415 & 0.433 & 0.432 & 3.379 & 1.338 \\ \cline{2-24} 
    \multicolumn{1}{c|}{} & Avg & 0.314 & 0.348 & \textbf{0.293} & \textbf{0.334} & 0.321 & 0.356 & 0.340 & 0.373 & 0.294 & 0.339 & \textbf{\underline{0.285}} & \textbf{\underline{0.328}} & 0.291 & 0.333 & 0.350 & 0.401 & 0.305 & 0.349 & 0.327 & 0.371 & 1.410 & 0.810 \\
    \hline \hline
    \multicolumn{1}{c|}{\multirow{5}{*}{\rotatebox{90}{ETTh1}}} & 96 & \textbf{0.384} & \textbf{\underline{0.395}} & 0.397 & 0.418 & 0.449 & 0.424 & 0.409 & 0.403 & 0.398 & 0.424 & 0.404 & 0.413 & 0.384 & 0.402 & 0.386 & 0.400 & \textbf{\underline{0.376}} & 0.419 & 0.449 & 0.459 & 0.865 & 0.713 \\
    \multicolumn{1}{c|}{} & 192 & \textbf{0.424} & \textbf{\underline{0.416}} & 0.434 & 0.439 & 0.503 & 0.453 & 0.467 & 0.444 & 0.449 & 0.427 & 0.454 & 0.440 & 0.436 & 0.429 & 0.437 & 0.432 & \textbf{\underline{0.420}} & 0.448 & 0.500 & 0.482 & 1.008 & 0.792 \\
    \multicolumn{1}{c|}{} & 336 & \textbf{\underline{0.451}} & \textbf{\underline{0.440}} & 0.468 & 0.457 & 0.540 & 0.477 & 0.509 & 0.472 & 0.492 & 0.466 & 0.497 & 0.462 & 0.491 & 0.469 & 0.481 & 0.459 & 0.459 & 0.465 & 0.521 & 0.496 & 1.107 & 0.809 \\
    \multicolumn{1}{c|}{} & 720 & 0.476 & \textbf{\underline{0.463}} & \textbf{\underline{0.469}} & 0.477 & 0.515 & 0.489 & 0.503 & 0.485 & 0.487 & 0.483 & 0.496 & 0.481 & 0.521 & 0.500 & 0.519 & 0.516 & 0.506 & 0.507 & 0.514 & 0.512 & 1.181 & 0.865  \\ \cline{2-24} 
    \multicolumn{1}{c|}{} & Avg & \textbf{\underline{0.434}} & \textbf{\underline{0.430}} & 0.442 & 0.448 & 0.502 & 0.461 & 0.472 & 0.451 & 0.457 & 0.450 & 0.463 & 0.449 & 0.458 & 0.450 & 0.456 & 0.452 & 0.440 & 0.460 & 0.496 & 0.487 & 1.040 & 0.795 \\
    \hline \hline
    \multicolumn{1}{c|}{\multirow{5}{*}{\rotatebox{90}{ETTh2}}} & 96 & \textbf{\underline{0.255}} & \textbf{\underline{0.320}} & 0.296 & 0.345 & 0.303 & 0.349 & 0.314 & 0.361 & 0.312 & 0.360 & 0.312 & 0.358 & 0.340 & 0.374 & 0.333 & 0.387 & 0.358 & 0.397 & 0.346 & 0.388 & 3.755 & 1.525 \\
    \multicolumn{1}{c|}{} & 192 & \textbf{\underline{0.316}} & \textbf{\underline{0.359}} & 0.374 & 0.394 & 0.391 & 0.399 & 0.407 & 0.411 & 0.387 & 0.405 & 0.397 & 0.408 & 0.402 & 0.414 & 0.477 & 0.476 & 0.429 & 0.439 & 0.456 & 0.452 & 5.602 & 1.931 \\
    \multicolumn{1}{c|}{} & 336 & \textbf{\underline{0.358}} & \textbf{\underline{0.394}} & 0.415 & 0.427 & 0.422 & 0.428 & 0.437 & 0.443 & 0.424 & 0.437 & 0.435 & 0.440 & 0.452 & 0.452 & 0.594 & 0.541 & 0.496 & 0.487 & 0.482 & 0.486 & 4.721 & 1.835 \\
    \multicolumn{1}{c|}{} & 720 & \textbf{\underline{0.419}} & \textbf{\underline{0.436}} & 0.425 & 0.444 & 0.429 & 0.449 & 0.434 & 0.448 & 0.433 & 0.453 & 0.436 & 0.449 & 0.462 & 0.468 & 0.831 & 0.657 & 0.463 & 0.474 & 0.515 & 0.511 & 3.647 & 1.625 \\ \cline{2-24} 
    \multicolumn{1}{c|}{} & Avg & \textbf{\underline{0.337}} & \textbf{\underline{0.377}} & 0.378 & 0.403 & 0.386 & 0.406 & 0.398 & 0.416 & 0.389 & 0.414 & 0.395 & 0.414 & 0.414 & 0.427 & 0.559 & 0.515 & 0.437 & 0.449 & 0.450 & 0.459 & 4.431 & 1.729 \\
    \hline \hline
    \multicolumn{1}{c|}{\multirow{5}{*}{\rotatebox{90}{Electricity}}} & 96 & 0.200 & 0.298 & \textbf{0.196} & \textbf{0.287} & 0.232 & 0.321 & 0.198 & 0.290 & 0.197 & 0.290 & 0.186 & \textbf{\underline{0.269}} & \textbf{\underline{0.168}} & 0.272 & 0.197 & 0.282 & 0.193 & 0.308 & 0.201 & 0.317 & 0.274 & 0.368 \\
    \multicolumn{1}{c|}{} & 192 & 0.205 & 0.303 & \textbf{0.199} & \textbf{0.291} & 0.234 & 0.325 & 0.202 & 0.293 & 0.201 & 0.292 & 0.190 & \textbf{\underline{0.273}} & \textbf{\underline{0.184}} & 0.289 & 0.196 & 0.285 & 0.201 & 0.315 & 0.222 & 0.334 & 0.296 & 0.386 \\
    \multicolumn{1}{c|}{} & 336 & 0.229 & 0.321 & \textbf{0.214} & \textbf{0.305} & 0.249 & 0.338 & 0.223 & 0.318 & 0.217 & 0.309 & 0.206 & \textbf{\underline{0.290}} & \textbf{\underline{0.198}} & 0.300 & 0.209 & 0.301 & 0.214 & 0.329 & 0.231 & 0.338 & 0.300 & 0.394 \\
    \multicolumn{1}{c|}{} & 720 & 0.268 & 0.353 & \textbf{0.254} & \textbf{0.335} & 0.289 & 0.366 & 0.259 & 0.341 & 0.253 & 0.339 & 0.247 & 0.322 & \textbf{\underline{0.220}} & \textbf{\underline{0.320}} & 0.245 & 0.333 & 0.246 & 0.355 & 0.254 & 0.361 & 0.373 & 0.439 \\ \cline{2-24} 
    \multicolumn{1}{c|}{} & Avg & 0.225 & 0.319 & \textbf{0.216} & \textbf{0.305} & 0.251 & 0.338 & 0.221 & 0.311 & 0.217 & 0.308 & 0.207 & \textbf{\underline{0.289}} & \textbf{\underline{0.192}} & 0.295 & 0.212 & 0.300 & 0.214 & 0.327 & 0.227 & 0.338 & 0.311 & 0.397 \\
    \hline \hline
    \multicolumn{1}{c|}{\multirow{5}{*}{\rotatebox{90}{Weather}}} & 96 & 0.202 & 0.251 & \textbf{\underline{0.171}} & \textbf{\underline{0.214}} & 0.212 & 0.251 & 0.213 & 0.260 & 0.203 & 0.244 & 0.177 & 0.218 & 0.172 & 0.220 & 0.196 & 0.255 & 0.217 & 0.296 & 0.266 & 0.336 & 0.300 & 0.384 \\
    \multicolumn{1}{c|}{} & 192 & 0.250 & 0.289 & \textbf{\underline{0.217}} & \textbf{\underline{0.254}} & 0.261 & 0.288 & 0.269 & 0.300 & 0.247 & 0.277 & 0.222 & 0.259 & 0.219 & 0.261 & 0.237 & 0.296 & 0.276 & 0.336 & 0.307 & 0.367 & 0.598 & 0.544 \\
    \multicolumn{1}{c|}{} & 336 & 0.299 & 0.321 & \textbf{\underline{0.274}} & \textbf{\underline{0.293}} & 0.313 & 0.324 & 0.330 & 0.341 & 0.297 & 0.311 & 0.277 & 0.297 & 0.280 & 0.306 & 0.283 & 0.335 & 0.339 & 0.380 & 0.359 & 0.395 & 0.578 & 0.523 \\
    \multicolumn{1}{c|}{} & 720 & 0.380 & 0.368 & \textbf{0.351} & \textbf{\underline{0.343}} & 0.386 & 0.372 & 0.404 & 0.389 & 0.368 & 0.356 & 0.352 & 0.347 & 0.365 & 0.359 & \textbf{\underline{0.345}} & 0.381 & 0.403 & 0.428 & 0.419 & 0.428 & 1.059 & 0.741 \\ \cline{2-24} 
    \multicolumn{1}{c|}{} & Avg & 0.283 & 0.307 & \textbf{\underline{0.253}} & \textbf{\underline{0.276}} & 0.293 & 0.309 & 0.304 & 0.323 & 0.279 & 0.297 & 0.257 & 0.280 & 0.259 & 0.287 & 0.265 & 0.317 & 0.309 & 0.360 & 0.338 & 0.382 & 0.634 & 0.548 \\
    \hline \hline
    \multicolumn{1}{c|}{\multirow{5}{*}{\rotatebox{90}{Exchange}}} & 96 & 0.131 & 0.269 & \textbf{\underline{0.086}} & \textbf{\underline{0.209}} & 0.142 & 0.261 & 0.137 & 0.260 & 0.091 & 0.212 & 0.109 & 0.236 & 0.107 & 0.234 & 0.088 & 0.218 & 0.148 & 0.278 & 0.197 & 0.323 & 0.847 & 0.752 \\
    \multicolumn{1}{c|}{} & 192 & 0.208 & 0.344 & \textbf{\underline{0.174}} & \textbf{\underline{0.299}} & 0.224 & 0.339 & 0.222 & 0.341 & 0.183 & 0.304 & 0.205 & 0.327 & 0.226 & 0.344 & 0.176 & 0.315 & 0.271 & 0.380 & 0.300 & 0.369 & 1.204 & 0.895 \\
    \multicolumn{1}{c|}{} & 336 & \textbf{\underline{0.309}} & 0.436 & 0.319 & \textbf{\underline{0.408}} & 0.377 & 0.448 & 0.372 & 0.447 & 0.328 & 0.417 & 0.356 & 0.436 & 0.367 & 0.448 & 0.313 & 0.427 & 0.460 & 0.500 & 0.509 & 0.524 & 1.672 & 1.036 \\
    \multicolumn{1}{c|}{} & 720 & \textbf{\underline{0.476}} & \textbf{\underline{0.554}} & 0.875 & 0.701 & 0.939 & 0.736 & 0.912 & 0.727 & 0.880 & 0.704 & 0.888 & 0.716 & 0.964 & 0.746 & 0.839 & 0.695 & 1.195 & 0.841 & 1.447 & 0.941 & 2.478 & 1.310 \\ \cline{2-24} 
    \multicolumn{1}{c|}{} & Avg & \textbf{\underline{0.281}} & \textbf{\underline{0.401}} & 0.364 & 0.404 & 0.421 & 0.446 & 0.411 & 0.444 & 0.371 & 0.409 & 0.390 & 0.429 & 0.416 & 0.443 & 0.354 & 0.414 & 0.519 & 0.500 & 0.613 & 0.539 & 1.550 & 0.998 \\
    \hline
    \multicolumn{2}{c|}{$1^{\text{st}}$ Count} & \multicolumn{2}{c|}{\textbf{22}} & \multicolumn{2}{c|}{\textbf{22}} & \multicolumn{2}{c|}{0} & \multicolumn{2}{c||}{0} & \multicolumn{2}{c|}{2} & \multicolumn{2}{c|}{13} & \multicolumn{2}{c|}{8} & \multicolumn{2}{c|}{1} & \multicolumn{2}{c|}{2} & \multicolumn{2}{c|}{0} & \multicolumn{2}{c}{0} \\
    \shline
    \end{tabular}
    } 
    \label{tab:cross}
    \vspace{-0.5cm}
\end{table*}

\subsection{Implementation Details}
\label{Implementation}

All experiments are conducted using the diffusion network from~\citep{diffae}, which consists of 20 MLP layers, with each layer having a hidden dimension of 2048. The number of diffusion steps is set to 200. During the pre-training stage, the model is trained on a 40G A100 GPU. For long and short-term forecasting tasks, the batch size is set to 1024 for the ETT, Weather, Solar, and Exchange datasets, and 128 or less for the remaining datasets, with a uniform iteration count of 100k. For cross-domain training, the batch size is set to 32, and the iteration count is 500k. All experiments use the Adam~\citep{kingma2014adam} optimizer with a learning rate of 10e-4 and the EMA rate of 0.9999 to optimize the parameters~\citep{dhariwal2021diffusion}. During the pre-training phase, the input length equals the historical length plus the maximum future length. For instance, for tasks with historical length 512 and different prediction horizons $H=\{96, 192, 336, 720\}$, the input length to the pre-trained model is 512+720=1232. During testing, given 512 lengths of historical data, the model directly generates future data with a maximum length of 720. For items shorter than 720, slicing is employed for comparison. Notable that due to the DropLast~\citep{paszke2019pytorch} setting in the code, different batch sizes among methods lead to different testset lengths. We strictly ensure that the length of testsets in the comparison is consistent.

\begin{table}[t]
\begin{center}
\caption{\textbf{Long-term forecasting results}. All results are averaged from four different forecasting horizons $\{96,192,336,720\}$. \textcolor{red}{Red} and \textcolor{blue}{blue} indicates the best and the second best performance.}
\label{tab:long-term-forecasting}
\begin{small}
\scalebox{0.68}{
\setlength \tabcolsep{3pt}
\begin{tabular}{c|cc|cc|cc|cc|cc|cc|cc|cc|cc|cc}
\toprule
Methods &\multicolumn{2}{c|}{Ours}&\multicolumn{2}{c|}{TimeLLM}&\multicolumn{2}{c|}{LLM4TS}&\multicolumn{2}{c|}{GPT4TS}&\multicolumn{2}{c|}{DLinear}&\multicolumn{2}{c|}{PatchTST}&\multicolumn{2}{c|}{FEDformer}&\multicolumn{2}{c|}{Autoformer}&\multicolumn{2}{c|}{Informer}&\multicolumn{2}{c}{Pyraformer}\\
\midrule
Metric &  MSE & MAE & MSE & MAE & MSE & MAE & MSE & MAE & MSE & MAE & MSE & MAE & MSE & MAE& MSE & MAE & MSE & MAE & MSE & MAE \\
\midrule
ETTh1& \textcolor{red}{0.391} & \textcolor{blue}{0.423} & 0.408 & 0.423 & \textcolor{blue}{0.404} & \textcolor{red}{0.418} & 0.428 & 0.426 & 0.423 & 0.438 & 0.413 & 0.461 & 0.440 & 0.677 & 0.496 & 0.487 & 1.040 & 0.795 & 0.827 & 0.703\\
\midrule
ETTh2& \textcolor{red}{0.300} & \textcolor{red}{0.365} & 0.334 & 0.383 & \textcolor{blue}{0.331} & \textcolor{blue}{0.383} & 0.355 & 0.395 & 0.431 & 0.447 & 0.370 & 0.379 & 0.437 & 0.449 & 0.450 & 0.459 & 4.431 & 1.729 & 0.829 & 0.703 \\
\midrule
ETTm1 & 0.349 & \textcolor{blue}{0.377} & \textcolor{red}{0.329} & \textcolor{red}{0.372} & \textcolor{blue}{0.343} & 0.378 & 0.352 & 0.383 & 0.357 & 0.379 & 0.351 & 0.381 & 0.448 & 0.452 & 0.588 & 0.517 & 0.961 & 0.734 & 0.691 & 0.607 \\
\midrule
ETTm2 & 0.262 & 0.314 & \textcolor{blue}{0.251} & \textcolor{blue}{0.313} & \textcolor{red}{0.251} & \textcolor{red}{0.313} & 0.267 & 0.326 & 0.267 & 0.334 & 0.255 & 0.315 & 0.305 & 0.349 & 0.327 & 0.371 & 1.410 & 0.810 & 1.498 & 0.869 \\
\midrule
Weather & 0.234 & 0.289 & \textcolor{red}{0.225} & \textcolor{red}{0.257} & \textcolor{blue}{0.223} & \textcolor{blue}{0.260} & 0.237 & 0.271 & 0.249 & 0.300 & 0.226 & 0.264 & 0.310 & 0.360 & 0.338 & 0.382 & 0.634 & 0.521 & 0.815 & 0.717 \\
\midrule
Electricity & 0.165 & 0.273 & \textcolor{red}{0.158} & \textcolor{red}{0.252} & \textcolor{blue}{0.159} & \textcolor{blue}{0.253} & 0.167 & 0.263 & 0.166 & 0.264 & 0.164 & 0.253 & 0.214 & 0.327 & 0.227 & 0.338 & 0.311 & 0.397 & 0.382 & 0.445\\
\midrule
Traffic & 0.393 & 0.279 & \textcolor{red}{0.388} & \textcolor{red}{0.264} & 0.401 & 0.273 & 0.414 & 0.295 & 0.437 & 0.264 & \textcolor{blue}{0.391} & \textcolor{blue}{0.264} & 0.611 & 0.376 & 0.628 & 0.379 & 0.764 & 0.416 & 1.176 & 0.469\\
\bottomrule
\end{tabular}
}
\end{small}
\end{center}
\end{table}

\begin{table}[t]
\vspace{-0.5cm}
\begin{center}
\caption{\textbf{Zero-shot forecasting results}. All results are averaged from four different forecasting horizons $\{96,192,336,720\}$. \textcolor{red}{Red} and \textcolor{blue}{blue} indicates the best and the second best performance.}
\label{tab:zero-shot-forecasting}
\begin{small}
\scalebox{0.78}{
\setlength \tabcolsep{3pt}
\begin{tabular}{c|cc|cc|cc|cc|cc|cc|cc|cc}
\toprule
Methods&\multicolumn{2}{c|}{Ours}&\multicolumn{2}{c|}{TimeLLM}&\multicolumn{2}{c|}{LLMTime}&\multicolumn{2}{c|}{GPT4TS}&\multicolumn{2}{c|}{DLinear}&\multicolumn{2}{c|}{PatchTST}&\multicolumn{2}{c|}{TimesNet}&\multicolumn{2}{c}{Autoformer}\\
\midrule
Metric & MSE & MAE & MSE & MAE & MSE & MAE & MSE & MAE & MSE & MAE & MSE & MAE& MSE & MAE & MSE & MAE \\
\midrule
{ETTh1} $\rightarrow$ \rotatebox{0}{ETTh2} & \textcolor{red}{0.303} & \textcolor{red}{0.368} & \textcolor{blue}{0.353} & \textcolor{blue}{0.387} &0.992 &0.708 & 0.406 & 0.422 & 0.493 & 0.488 & 0.380 & 0.405 & 0.421 & 0.431 & 0.582 & 0.548 \\
\midrule
{ETTh1} $\rightarrow$ \rotatebox{0}{ETTm2} & \textcolor{blue}{0.311} & 0.367 & \textcolor{red}{0.273} & \textcolor{red}{0.340} & 1.867 &0.869 & 0.325 & 0.363 & 0.415 & 0.452 & 0.314 & \textcolor{blue}{0.360} & 0.327 & 0.361 & 0.457 & 0.483 \\
\midrule
{ETTh2} $\rightarrow$ \rotatebox{0}{ETTh1} & \textcolor{red}{0.446} & \textcolor{red}{0.456} & \textcolor{blue}{0.479} & \textcolor{blue}{0.474} &1.961 & 0.981 & 0.757 & 0.578 & 0.703 & 0.574 & 0.565 & 0.513 & 0.865 & 0.621 & 0.757 & 0.608 \\
\midrule
{ETTh2} $\rightarrow$ \rotatebox{0}{ETTm2} & \textcolor{blue}{0.316} & \textcolor{blue}{0.364} & \textcolor{red}{0.272} & \textcolor{red}{0.341} &1.867 &0.869 & 0.335 & 0.370 & 0.328 & 0.386 & 0.325 & 0.365 & 0.342 & 0.376 & 0.366 & 0.411 \\
\midrule
{ETTm1} $\rightarrow$ \rotatebox{0}{ETTh2}  & \textcolor{red}{0.333} & \textcolor{red}{0.389} & \textcolor{blue}{0.381} & \textcolor{blue}{0.412} &0.992 &0.708 & 0.433 & 0.439 & 0.464 & 0.475 & 0.439 & 0.438 & 0.457 & 0.454 & 0.470 & 0.479 \\
\midrule
{ETTm1} $\rightarrow$ \rotatebox{0}{ETTm2}  & \textcolor{red}{0.266} & \textcolor{red}{0.317} & \textcolor{blue}{0.268} & \textcolor{blue}{0.320} &1.867 &0.869 & 0.313 & 0.348 & 0.335 & 0.389 & 0.296 & 0.334 & 0.322 & 0.354 & 0.469 & 0.484 \\
\midrule
{ETTm2} $\rightarrow$ \rotatebox{0}{ETTh2}  & \textcolor{red}{0.311} & \textcolor{red}{0.369} & \textcolor{blue}{0.354} & \textcolor{blue}{0.400} &0.992 &0.708 & 0.435 & 0.443 & 0.455 & 0.471 & 0.409 & 0.425 & 0.435 & 0.443 & 0.423 & 0.439 \\
\midrule
{ETTm2} $\rightarrow$ \rotatebox{0}{ETTm1} & \textcolor{red}{0.384} & \textcolor{red}{0.404} & \textcolor{blue}{0.414} & \textcolor{blue}{0.438} &1.933 &0.984 & 0.769 & 0.567 & 0.649 & 0.537 & 0.568 & 0.492 & 0.769 & 0.567 & 0.755 & 0.591 \\
\bottomrule
\end{tabular}
}
\end{small}
\end{center}
\vspace{-0.5cm}
\end{table}

\subsection{Cross-Domain Time Series Forecasting}

The cross-domain forecasting follows the experimental setup of UniTime~\citep{unitime}. Table \ref{tab:cross} is copied from UniTime and presents the overall performance. Two vertical lines are used to demarcate the table. The left side shows the results of cross-domain training, with forecasting horizons slices at the maximum prediction length. The data on the right side comes from the officially provided results of training models separately for each specific length and dataset. Our method achieves comparable performance to the SOTA UniTime. Notably, UniTime introduces domain priors for cross-domain modeling, i.e., distinguishing different domains during training and testing based on the text descriptions of each dataset. To ensure experimental integrity and explore the diffusion model's ability to model complex distributions, we do not introduce conditions for distinguishing domains. In comparison with the SOTA LLMs-based paradigm GPT4TS~\citep{gpt4ts} and deep models paradigm PatchTST~\citep{patchtst} (both without introducing domain priors), GPD outperforms them comprehensively, suggesting the potential of the GPD paradigm as a fundamental and unified model from the side.

\subsection{Long and Short-Term Time Series Forecasting}

\textbf{Long-range}: Following the settings and results for long-range forecasting in TimeLLM~\citep{timellm} and LLM4TS~\citep{llm4ts}, we pretrain the model for specific datasets. Table \ref{tab:long-term-forecasting} reports the average performance comparison of our method with LLMs-based and deep models paradigm~\citep{fedformer,autoformer,informer,pyraformer} in different horizons. Overall quantification is provided in Appendix.

\begin{figure}[t]
    \centering
    \begin{minipage}[t]{0.66\textwidth}
    \vspace{0pt}
        \centering
        \small
          \begin{tabular}{cccccc}
            \toprule
            & Solar & Electricity & Traffic & Taxi & Wikipedia\\
            \midrule
            Method & \multicolumn{5}{c}{MSE} \\
            \midrule
             TimeGrad & 9.9e2 & 2.1e5 & 4.6e-4 & 2.4e & 3.1e7 \\
             CSDI & 9.4e2 & 2.4e5 & 4.4e-4 & - & - \\
             SSSD & 5.4e2  & 2.3e5 & 4.5e-4 & 2.3e & 2.99e7 \\
             $\text{D}^3\text{VAE}$ & 9.2e2 & 2.4e5 & 4.5e-4 & 2.4e & 3.2e7 \\
             LDT & 7.7e2 & 1.6e5 & 4.1e-4 & 2.2e & 2.92e7 \\
             Ours & \textbf{4.1e2} & \textbf{1.4e5} & \textbf{3.3e-4} & \textbf{1.9e} & \textbf{2.32e7} \\
            \bottomrule
          \end{tabular}
          \captionof{table}{Quantitative comparison of different diffusion-based methods in short-term forecasting}
          \label{tab:short-term-forecasting}
    \end{minipage}
    \hfill
    \begin{minipage}[t]{0.33\textwidth}
    \vspace{0pt}
        \centering
        \includegraphics[width=\linewidth]{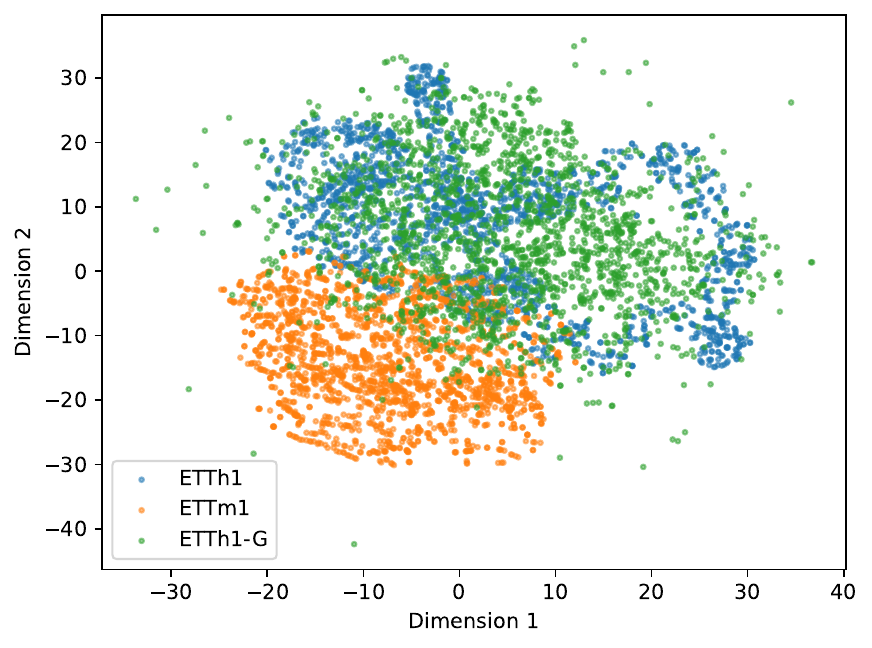}
        \caption{t-SNE Visualization of GPD modeling capability.}
        \label{fig:SNE}
    \end{minipage}
\vspace{-0.7cm}
\end{figure}

\textbf{Short-range}: Due to the current absence of diffusion-based methods in long-range forecasting, we follow the settings and results of LDT~\citep{ldt} for short-term forecasting and compare with similar works~\citep{timegrad,csdi,sssd,d3vae}. Table \ref{tab:short-term-forecasting} reports that our method outperforms the SOTA diffusion paradigm.

\subsection{Zero-Shot Forecasting}

Zero-shot forecasting is defined as a model trained on dataset \textit{A} being used to predict an unseen dataset \textit{B}. Following the experiments in TimeLLM, we perform a comprehensive zero-shot learning on the ETT dataset, as shown in Table \ref{tab:zero-shot-forecasting}. Based on long-term forecasting, we discover an interesting phenomenon. Under traditional mechanisms, although we are slightly inferior to TimeLLM, our zero-shot learning capability significantly surpasses that of TimeLLM \citep{timellm}. Regression models primarily learn the mapping between historical and future data within the dataset. In the context of zero-shot learning, the mapping learned by the model on dataset \textit{A} may lead to suboptimal generalization performance on dataset \textit{B}. In contrast, GPD paradigms model the underlying patterns and structures of time series data, thereby providing rich prior information that facilitates superior generalization.

\section{Model Analysis}

In this section, we conduct numerous analysis experiments on model pre-training and diffusion sampling to explore the impact of various components on the performance of zero-shot TSF. In Appendix \ref{imputation} and \ref{classfication}, we additionally demonstrate the zero-shot imputation and classification capabilities of proposed GPD paradigm.

\textbf{Distribution Matching}: We utilize a diffusion model pre-trained on the ETTh1 dataset to randomly generate 2000 samples. Subsequently, we randomly sample 2000 samples from both ETTh1 and ETTm1 datasets and employ t-SNE to visualize the data distribution. As depicted in Fig. \ref{fig:SNE}, GPD exhibits robust distribution modeling capability, which is crucial for downstream tasks. Unlike current cross-modal approaches based on LLMs that may introduce concept drift phenomena, GPD offers better interpretability for modeling temporal modalities.

\begin{figure}[t]
  \vspace{-0.4cm}
  \centering
  \begin{minipage}[t]{0.5\textwidth}
  \subfigure[ETTh1-$\epsilon_\theta$ Mode]{
      \includegraphics[width=0.48\textwidth]{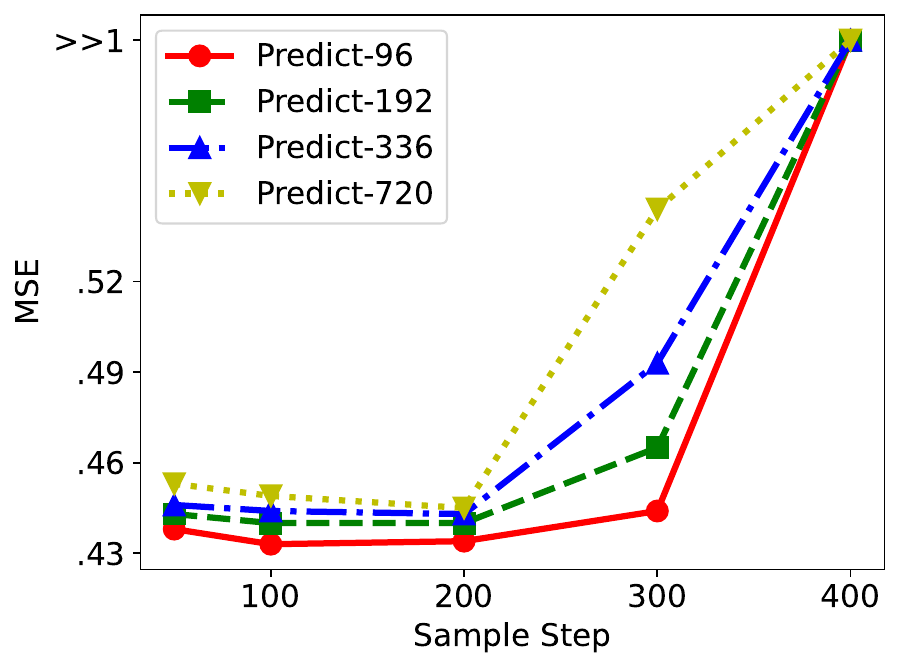}
      \hspace{-0.5em}
      \includegraphics[width=0.48\textwidth]{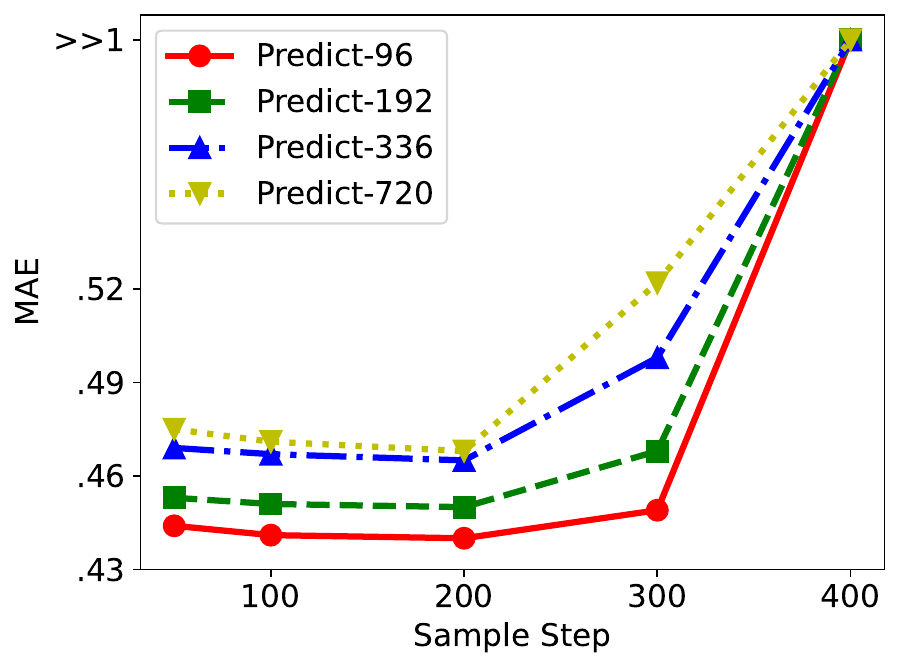}
  }
  \end{minipage}
\hspace{-0.5em}
  \begin{minipage}[t]{0.5\textwidth}
  \subfigure[Weather-$X_\theta$ Mode]{
      \includegraphics[width=0.49\textwidth]{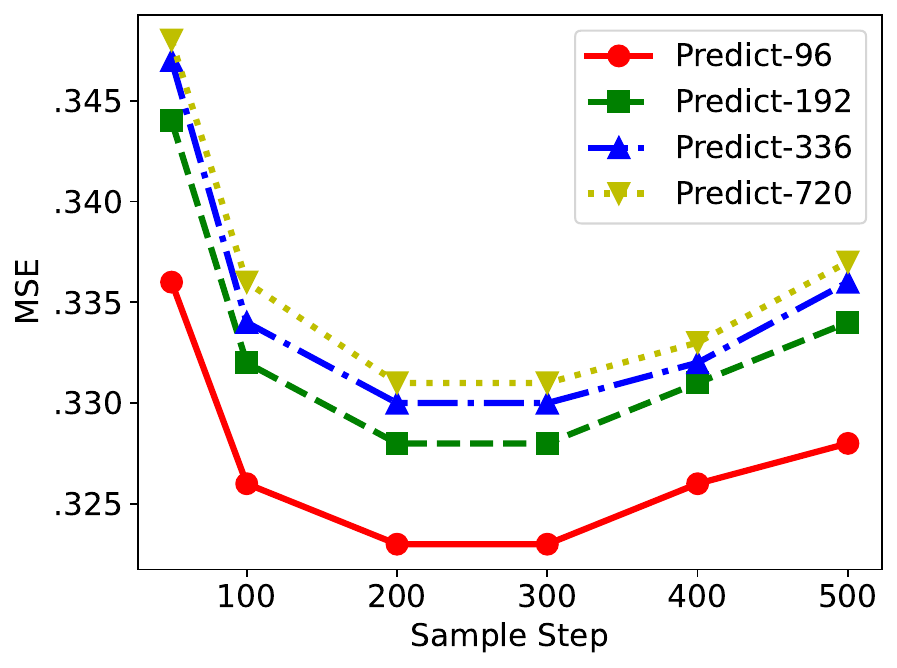} 
      \hspace{-0.5em}
      \includegraphics[width=0.48\textwidth]{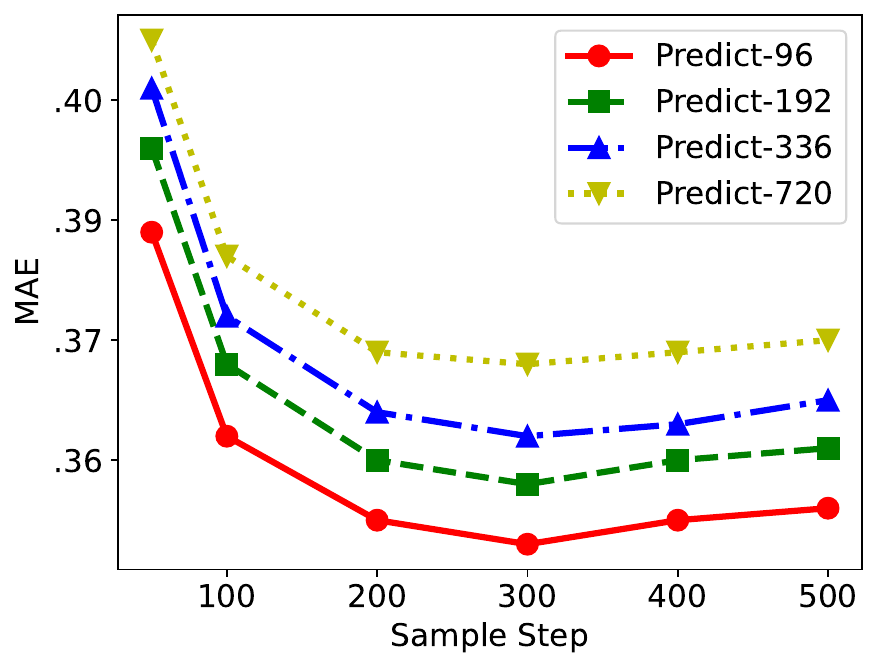}
  }
  \end{minipage}
  \vspace{-8pt}
  \caption{The impact of different sampling step settings and prediction strategies on forecasting performance. $>>1$ indicates that the metric is far greater than one.}
  \label{eps_mse_mae}
  \vspace{-0.5cm}
\end{figure}

\textbf{Sampling Steps and Prediction Strategy}: The number of diffusion sampling steps directly affects the model's inference time. We find that excessively high sampling steps may increase the transition error at each step, which is particularly evident in predicting $\epsilon_\theta$, as shown in Fig. \ref{eps_mse_mae}(a). Conversely, the $X_\theta$ prediction mode exhibits more stability. However, both modes indicate that excessive sampling steps do not improve model performance, hence a range between 50 to 200 is a reasonable choice. In addition, Table \ref{tab:mode} shows the comparison of different prediction models, and $X_\theta$ mode is usually better.

\begin{figure}[t]
    \centering
    \begin{minipage}[t]{0.675\textwidth}
    \vspace{0pt}
        \centering
        \subfigure{
            \includegraphics[width=0.48\linewidth]{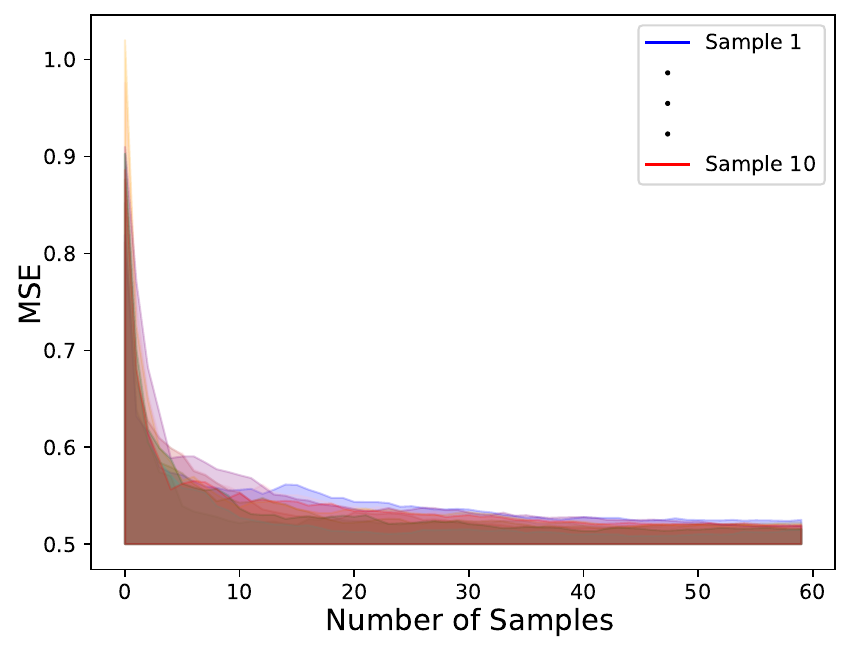}
        }
        \hspace{-0.8em}
        \subfigure{
            \includegraphics[width=0.487\linewidth]{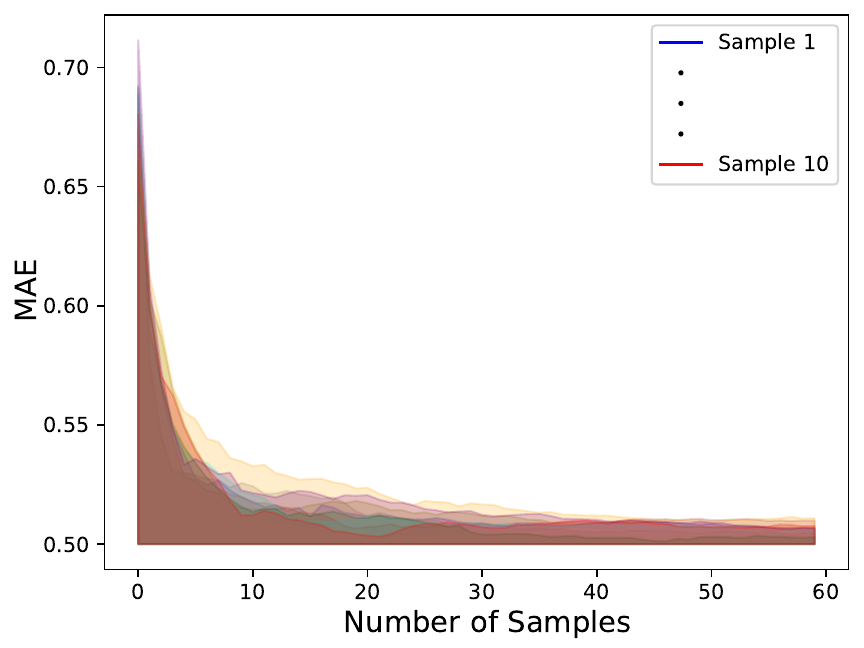}
        }
        \caption{\textbf{Analysis of Average Sampling Times.}}
        \label{ab_num_sample_mae}
    \end{minipage}
    \hfill
    \begin{minipage}[t]{0.319\textwidth}
    \vspace{3pt}
    \centering
    \includegraphics[height=0.80\textwidth]{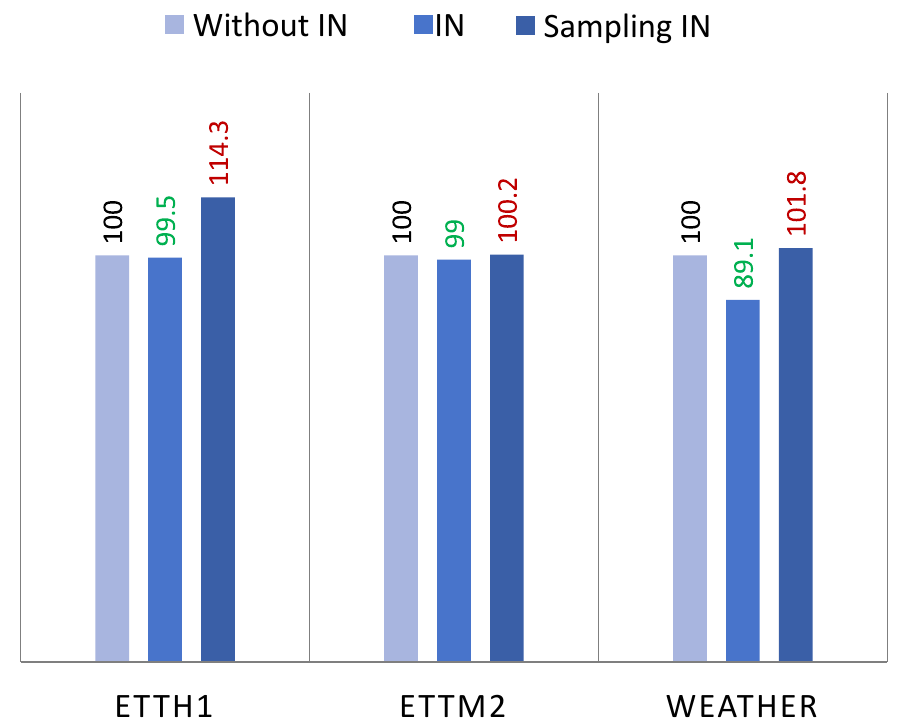}
    \caption{\textbf{Normalization Analysis.}}
    \label{IN}
    \end{minipage}
\end{figure}

\textbf{Sample Averaging}: Due to GPD's lack of fine-tuning the pre-trained model for forecasting tasks and relying entirely on prior knowledge from the pure generative model, results from single sampling exhibit significant randomness. As depicted in Fig. \ref{ab_num_sample_mae}, we explore the MSE and MAE of different samples with varying numbers of sampling for the same time series. It can be observed that with an increase in the number of sampling, the randomness bias tends to stabilize. When the sampling number exceed 5, the bias between different samples rapidly converges to a small range. To achieve better performance, we set a uniform sampling number of 50 in all experiments, where the performance bias does not exceed 5$\%$ compared to a sampling number of 5.

\textbf{Instance Normalization}: Instance normalization (IN), as demonstrated by RevIN~\citep{kim2021reversible} in many previous methods, effectively enhances forecasting performance. However, diffusion models are designed to model noisy data directly, making it challenging to incorporate IN into the denoiser. We introduce IN through data preprocessing. For instance, for input data of length $L$, we normalize the data using the mean and variance of sub-sequences of length $H$ or entire sequences. During sampling, we denormalize the predicted results using the mean and variance of the historical sequence. However, we find that the effectiveness of IN during the training process is marginal. Therefore, we propose sampling IN, which does not introduce IN during the training process but incorporates it during sampling. Across multiple datasets, sampling IN outperforms or is on par with using IN and not using IN, as shown in Fig. \ref{IN}. This is because incorporating historical sequences as prompts into the posterior distribution affects the pre-trained distribution. Sampling IN reduces the impact on the distribution while ensuring modeling of the original data distribution.

\begin{figure*}[t]
    \vspace{-0.5cm}
    \centering
    \subfigure[100 $\rightarrow$ 96]{\includegraphics[width=0.245\textwidth]{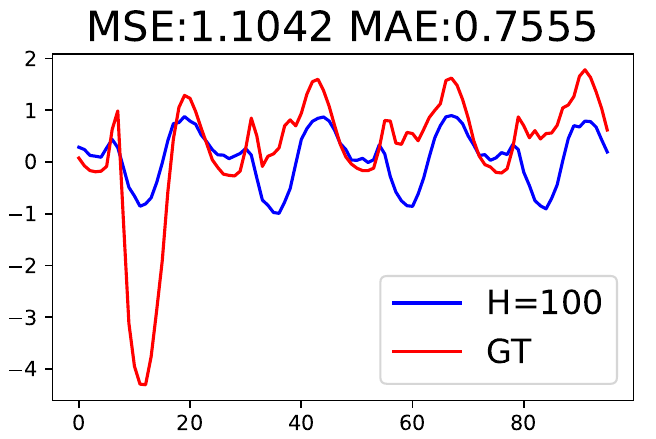}\label{fig: sub_figure1}}
    \subfigure[300 $\rightarrow$ 96]{\includegraphics[width=0.245\textwidth]{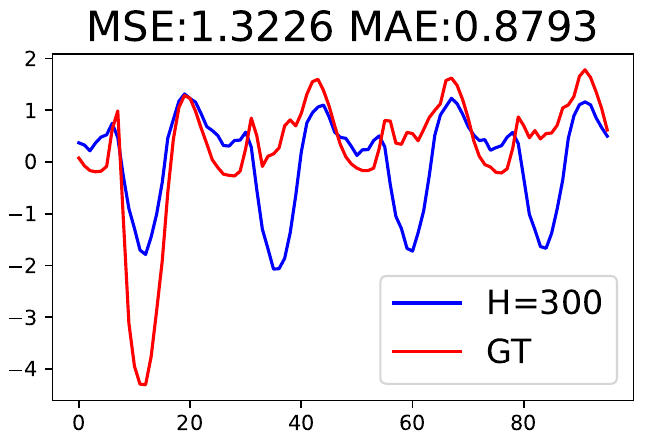}\label{fig: sub_figure2}}
    \subfigure[500 $\rightarrow$ 96]{\includegraphics[width=0.245\textwidth]{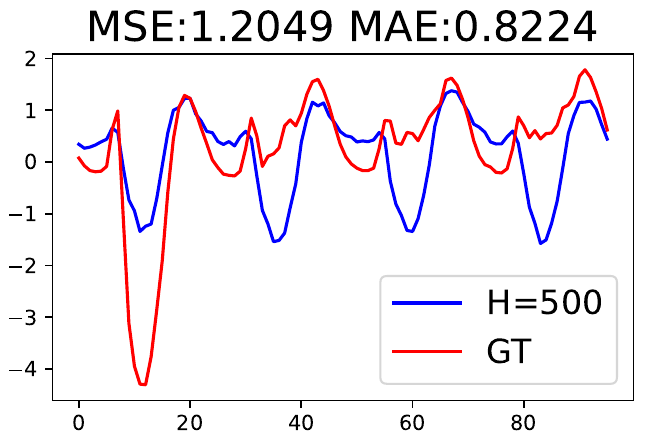}\label{fig: sub_figure3}}
    \subfigure[700 $\rightarrow$ 96]{\includegraphics[width=0.245\textwidth]{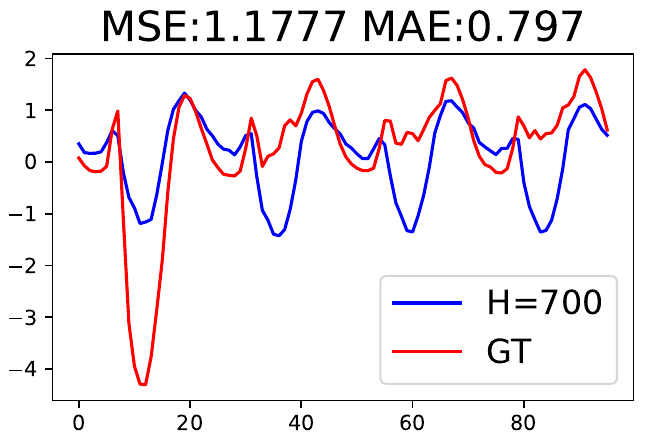}\label{fig: sub_figure4}}\vspace{-10pt}

    \subfigure[222 $\rightarrow$ 192]{\includegraphics[width=0.245\textwidth]{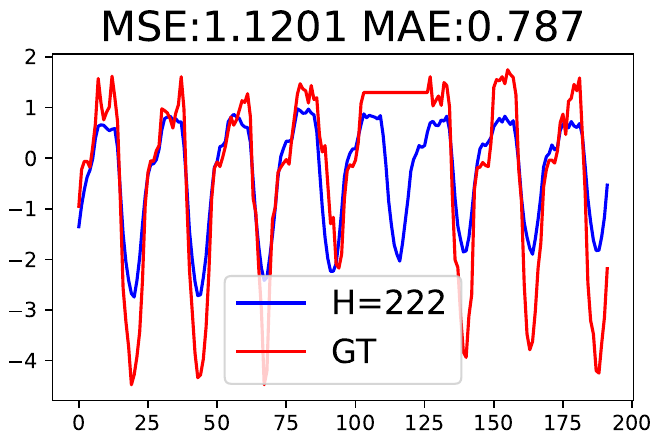}\label{fig: sub_figure5}}
    \subfigure[444 $\rightarrow$ 192]{\includegraphics[width=0.245\textwidth]{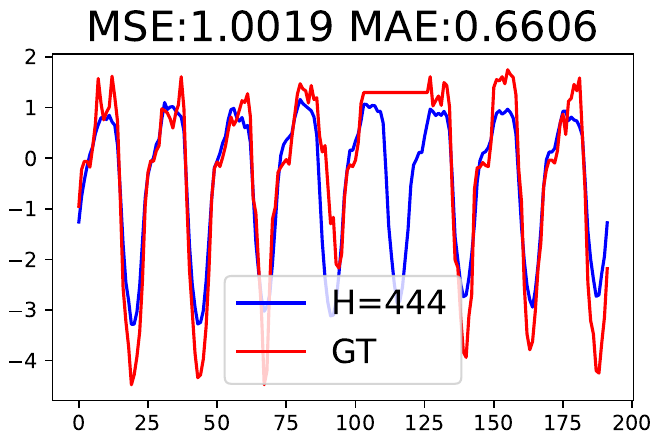}\label{fig: sub_figure6}}\hspace{6pt}
    \subfigure[50 $\rightarrow$ 1000]{\includegraphics[width=0.465\textwidth]{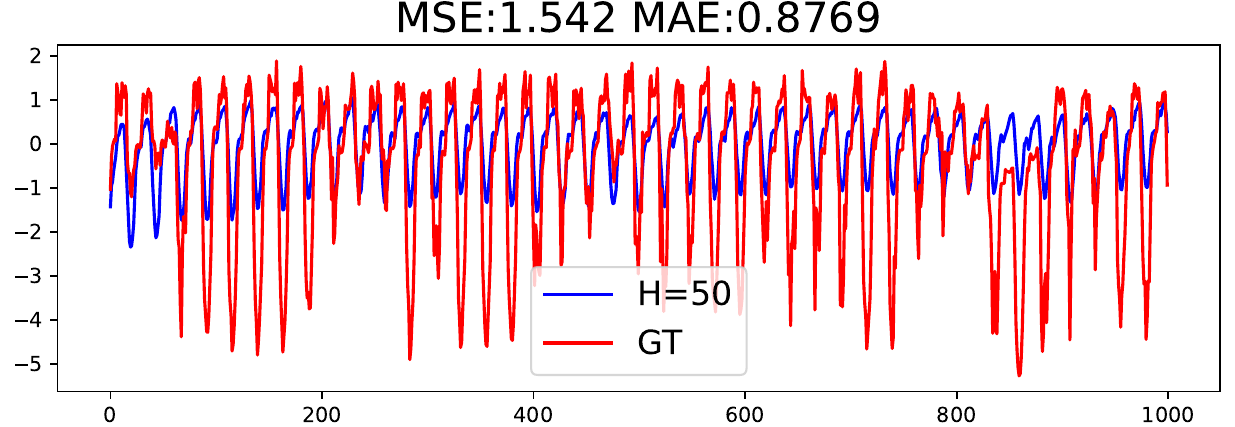}\label{fig: sub_figure7}}\hspace{4pt}
    \caption{\textbf{Visualization of Flexible Forecasting Results.}}
    \label{fig: flexible}
\vspace{-0.7cm}
\end{figure*}

\textbf{Flexible Forecasting of Arbitrary Lengths}: Addressing the complex forecasting needs of the real-world requires exceptional algorithmic flexibility. GPD demonstrates remarkable flexibility, which is unattainable in previous works. As shown in Fig. \ref{fig: flexible}, we visualize the forecasting results of an arbitrary-length future sequence $P$ from an arbitrary-length historical sequence $H$, simplified as $H \rightarrow P$. Notably, the above experiments are based on the same pre-trained model. In contrast, previous methods require training dedicated models separately for different lengths of $H$ and $P$.

\begin{table}[tbp]
\begin{center}
\caption{Quantitative comparison of different prediction modes in long-range forecasting task.}
\label{tab:mode}
\begin{small}
\setlength \tabcolsep{3pt}
\renewcommand{\arraystretch}{1.5}
\begin{tabular}{ccccccccccccccc}
\toprule
\multirow{2}{*}{Mode}&\multicolumn{2}{c}{ETTh1}&\multicolumn{2}{c}{ETTh2}&\multicolumn{2}{c}{ETTm1}&\multicolumn{2}{c}{ETTm2}&\multicolumn{2}{c}{Electricity}&\multicolumn{2}{c}{Weather}&\multicolumn{2}{c}{Traffic}\\
\cline{2-15}
& MSE & MAE & MSE & MAE & MSE & MAE & MSE & MAE & MSE & MAE & MSE & MAE & MSE & MAE\\
\midrule
$\epsilon_\theta$  & \textbf{0.391} & \textbf{0.423} & \textbf{0.300} & \textbf{0.365} & 0.375 & 0.417 & 0.296 & 0.341 & 0.176 & 0.275 & 0.237 & \textbf{0.280} & \textbf{0.393} & \textbf{0.279} \\
$X_\theta$         & 0.423 & 0.447 & 0.349 & 0.392 & \textbf{0.349} & \textbf{0.377} & \textbf{0.262} & \textbf{0.314} & \textbf{0.166} & \textbf{0.263} & \textbf{0.234} & 0.289 & 0.396 & 0.284 \\
\bottomrule
\end{tabular}
\end{small}
\end{center}
\end{table}

\section{Related Works}

\subsection{Large Language Models for Time Series}

Recently emerged works on LLMs for time series can be categorized into two pre-trained paradigms: Fine-tuning LLMs' powerful linguistic reasoning capabilities for time series modeling. And training a dedicated time series base model from scratch.

\textbf{Pre-trained LLMs}. GPT4TS~\citep{gpt4ts} pioneers the validation of the pre-trained paradigm's feasibility and universality. TEMPO~\citep{tempo} infuses more information into pre-trained LLM by decomposing the complex interactions between trend, seasonal, and residual components during fine-tuning. Additionally, numerous efforts are devoted to aligning natural language with time series. LLMTime~\citep{llmtime} actively aligns time series data with natural language to leverage LLMs' zero-shot logical inference capabilities. TimeLLM~\citep{timellm} reprograms pre-trained LLMs by adjusting the time series modality using text prompts. These methods preserve the integrity of LLMs. Furthermore, some works align modalities while fine-tuning to pursue better performance. $\text{S}^2\text{IP-LLM}$~\citep{s2ipllm} aligns the semantic space with time series embedding and predicts future based on prompts learned from the joint space. LLaTA~\citep{llata} addresses the non-aligned modality between pre-trained LLMs and time series via knowledge distillation.

\textbf{Time series foundation models}. Another attempt is to draw inspiration from foundation models in other fields, such as NLP~\citep{clip,gpt3} and CV~\citep{sora,sd}, to construct foundation models for time series. The key challenge lies in establishing robust mappings across sequential data from different feature domains. TimeGPT~\citep{garza2023timegpt} employs large-scale Transformer trained on massive datasets to enhance robustness. Lag-Llama~\citep{rasul2023lag} explores using lags as covariates in a pure decoder Transformer pre-trained on large corpora of diverse time series data across multiple domains. 

\subsection{Diffusion Models for Time Series}

TimeGrad~\citep{timegrad} first proposes combining DDPM and RNN for multivariate time series forecasting, recursively predicting in an autoregressive manner. ScoreGrad~\citep{yan2021scoregrad} is built on score matching, utilizing a diffusion model to assist an RNN-like module in step-by-step prediction. We acknowledge that diffusion models are also iterative models, and combining autoregressive and diffusion approaches often incurs high computational costs and accumulated errors \citep{wang2023exploiting}. To address this, DiffSTG~\citep{wen2023diffstg} combines spatial-temporal graphs and DDPM to propose a non-autoregressive framework, utilizing spatial-temporal graphs to guide the diffusion model in modeling spatial-temporal correlations. CSDI~\citep{csdi} and SSSD~\citep{sssd} introduce conditions into diffusion networks and train a conditional diffusion model with inductive bias losses. TimeDiff~\citep{shen2023non} models future data conditioned on historical data, then predicts future data in one shot through diffusion sampling. Most condition-guided methods rely on the initial prediction from the predictor, which is unfavorable for constructing fundamental and unified models. Additionally, some works attempt to leverage pre-trained diffusion paradigms. \citet{ldt} propose a Latent Diffusion Transformer (LDT) framework, a multivariate time series diffusion model that generates realistic multivariate time stamps guided by self-conditioning. Diffusion-TS~\citep{yuan2024diffusion} combines seasonal-trend decomposition and denoising sample reconstruction to build an unconditional model, then utilizes a classifier to guide conditional tasks. 

\section{Conclusion and Discussion}
\label{Conclusion}

\textbf{Conclusion}. We explore and propose a novel Generative Pre-trained Diffusion (GPD) paradigm for zero-shot TSF. In various real-world forecasting tasks, GPD demonstrates superior performance, surpassing methods based on LLMs and deep models for the first time. Compared to previous methods, our method is more general, versatile, and flexible. Extensive experiments and analyses validate the potential of the GPD paradigm and lay the foundation for future work.

\textbf{Discussion and Limitations}. Firstly, the current field of time series is not as thriving as the NLP or CV fields, lacking large-scale models and multi-modal datasets \citep{wahle2022identifying,srinivasan2021wit,schuhmann2021laion}, among other resources. This somewhat limits the diversity and development within the time series domain. Although we validate the effectiveness of the GPD paradigm, its generality requires further verification. For instance, the scaling ability, multi-modal integration, and controllability need to be explored further. Secondly, fine-tuning techniques for diffusion models, such as ControlNet~\citep{zhang2023adding} and Adapter~\citep{mou2024t2i,zhao2024uni}, have achieved significant success in other fields. Future exploration can focus on more downstream tasks. Furthermore, the use of simple baseline may led to sub-optimal performance, thus designing superior diffusion networks for time series is necessary. Lastly, similar to diffusion-based methods in other fields, our method does not have an advantage in speed compared to end-to-end models. We found that adopting acceleration strategies~\citep{ddim} seriously impacts performance, possibly due to the limited modeling capability of MLP or the lack of fine-tuning for downstream tasks.

\bibliography{refs}
\bibliographystyle{plainnat}

\clearpage
\appendix
\section{Expectation}
\label{Expectation}
We have validated the potential of the Generative Pre-trained Diffusion (GPD) paradigm as a zero-shot forecaster. In the Appendix, we further verify its capabilities in zero-shot imputation and classification. We have thoroughly explored the effectiveness and superiority of the GPD paradigm using a simple baseline network. Our aim is to lay the groundwork for foundation models in time series. In fact, we can pretrain a unified model, similar to cross-domain, to accomplish most downstream tasks. Extensive experiments demonstrate the feasibility of the GPD paradigm. We introduce many interesting points worthy of further follow-up and research. Code can be accessed from anonymous link \url{https://anonymous.4open.science/r/GPD}.

\section{Zero-Shot Imputation}
\label{imputation}
\begin{figure*}[htbp]
    \centering
    \subfigure[Missing 30\% Data]{\includegraphics[width=0.45\textwidth]{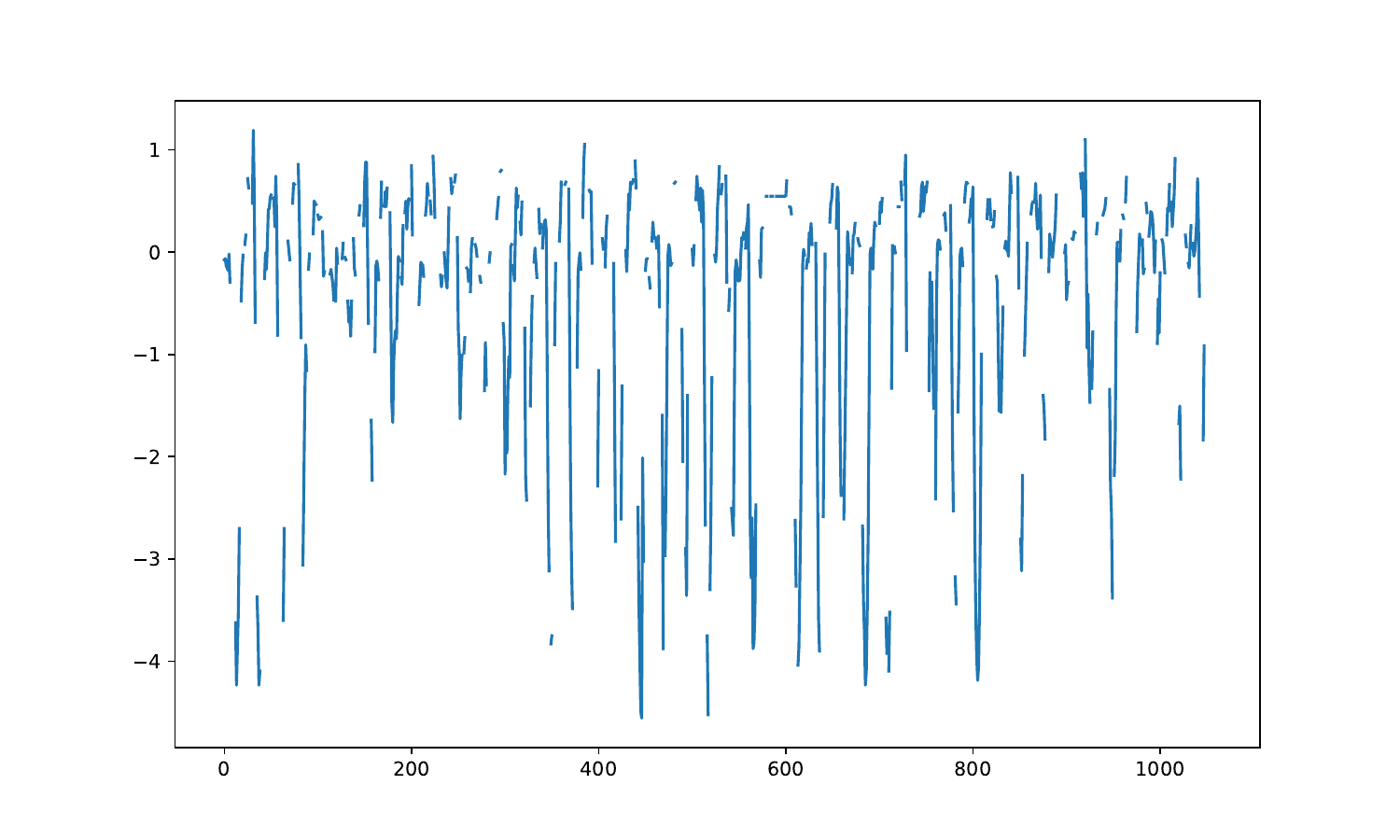}\label{fig: imputation30}}
    \subfigure[Imputation Result]{\includegraphics[width=0.45\textwidth]{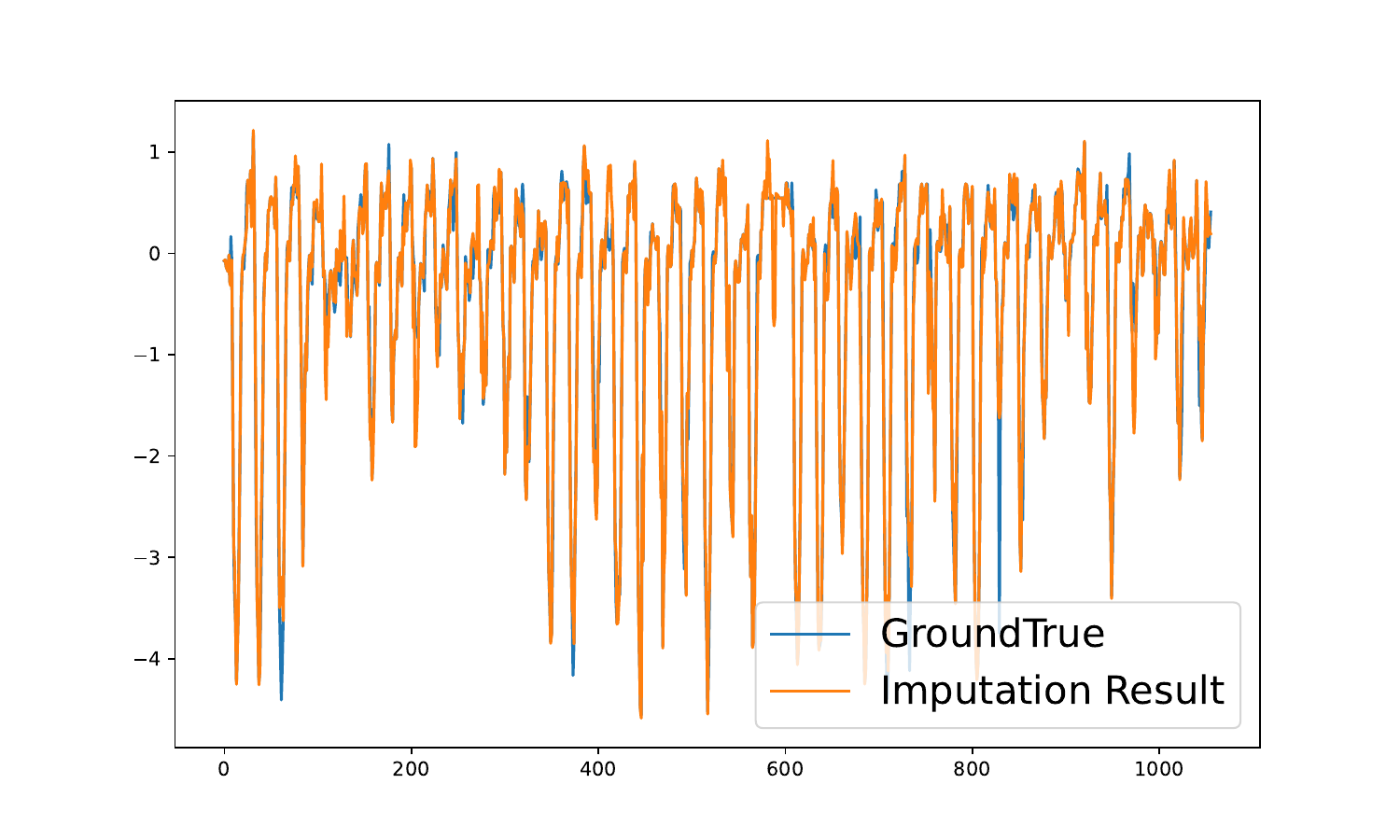}\label{fig: imputation_res30}}
    \vspace{-5pt}

    \subfigure[Missing 60\% Data]{\includegraphics[width=0.45\textwidth]{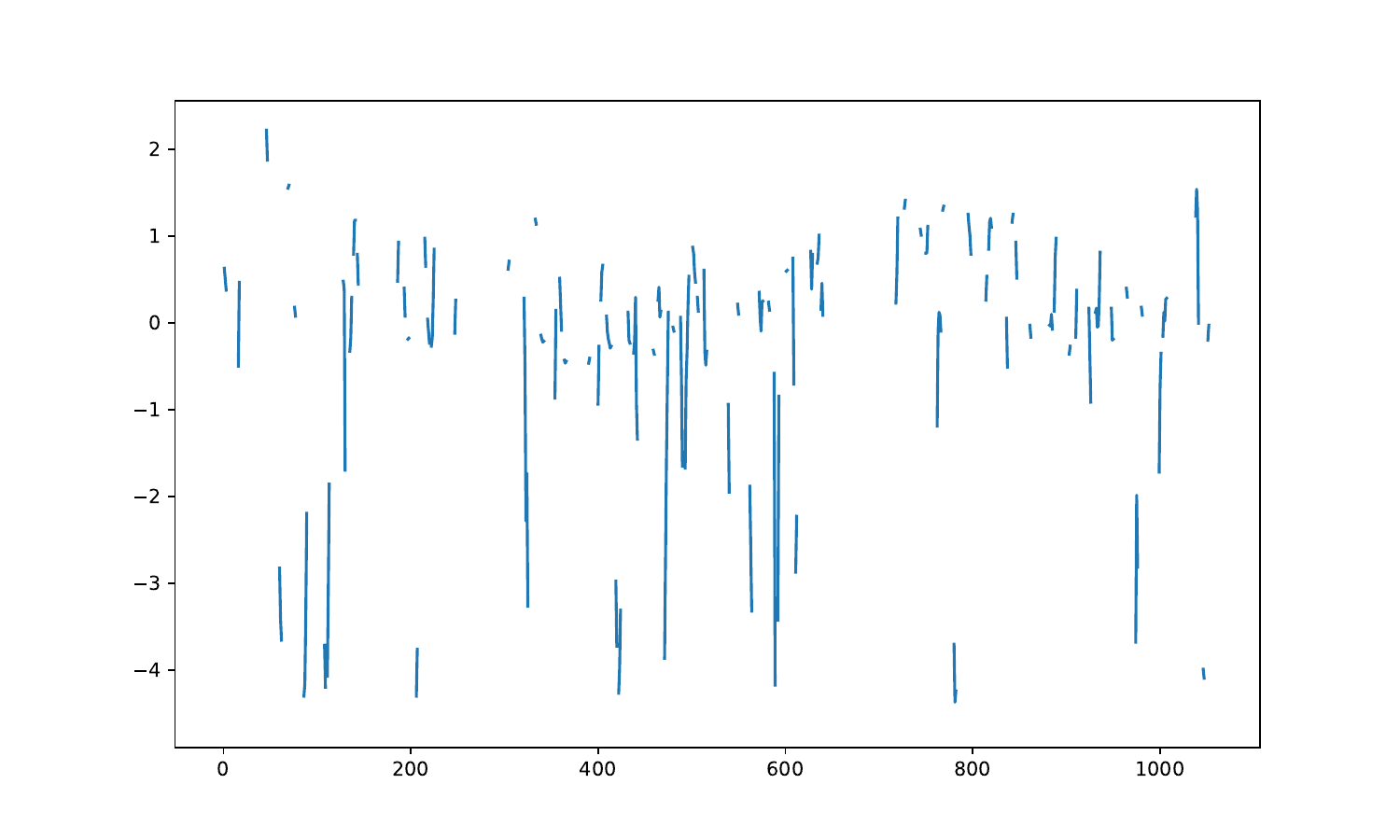}\label{fig: imputation60}}
    \subfigure[Imputation Result]{\includegraphics[width=0.45\textwidth]{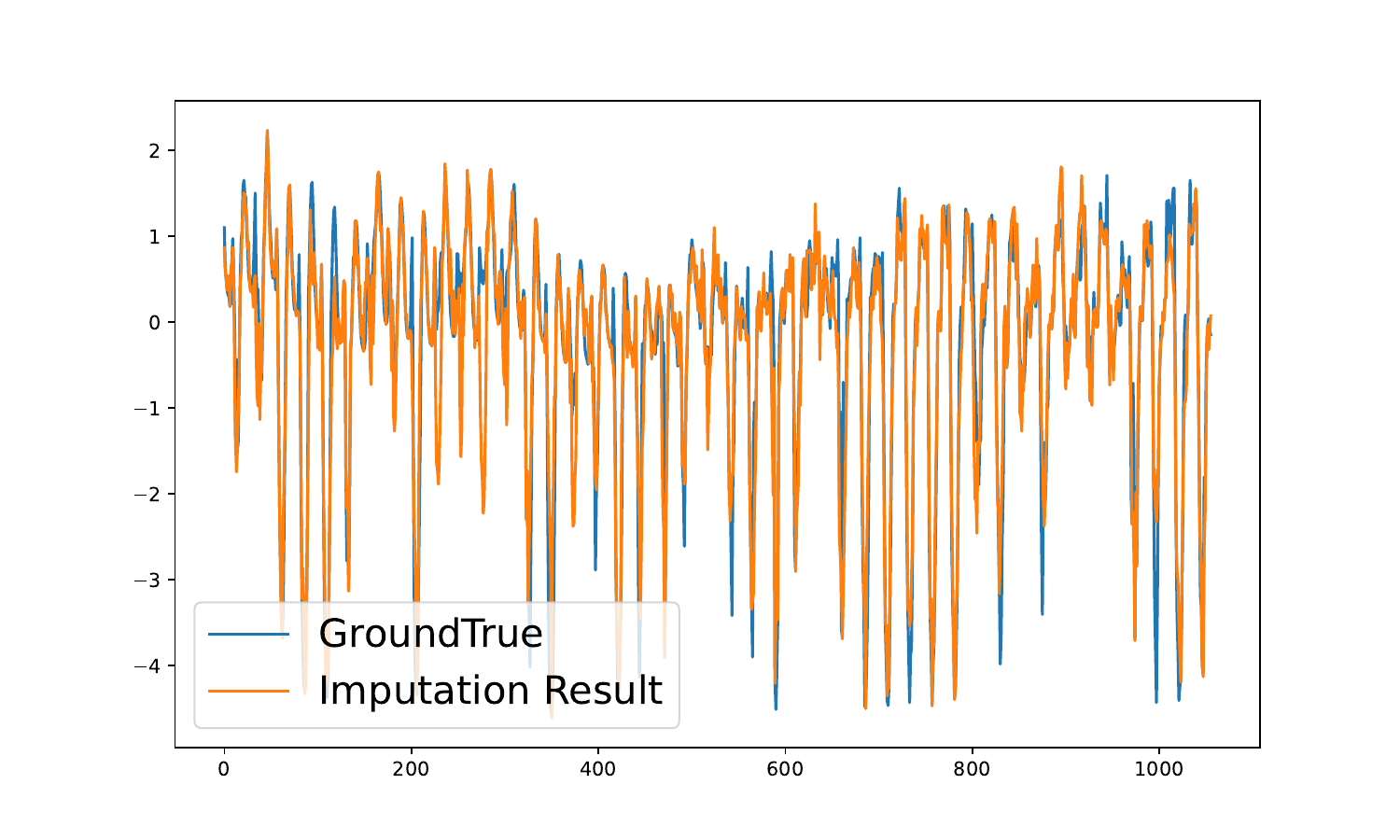}\label{fig: imputation_res60}}
    \vspace{-5pt}

    \subfigure[Missing 90\% Data]{\includegraphics[width=0.45\textwidth]{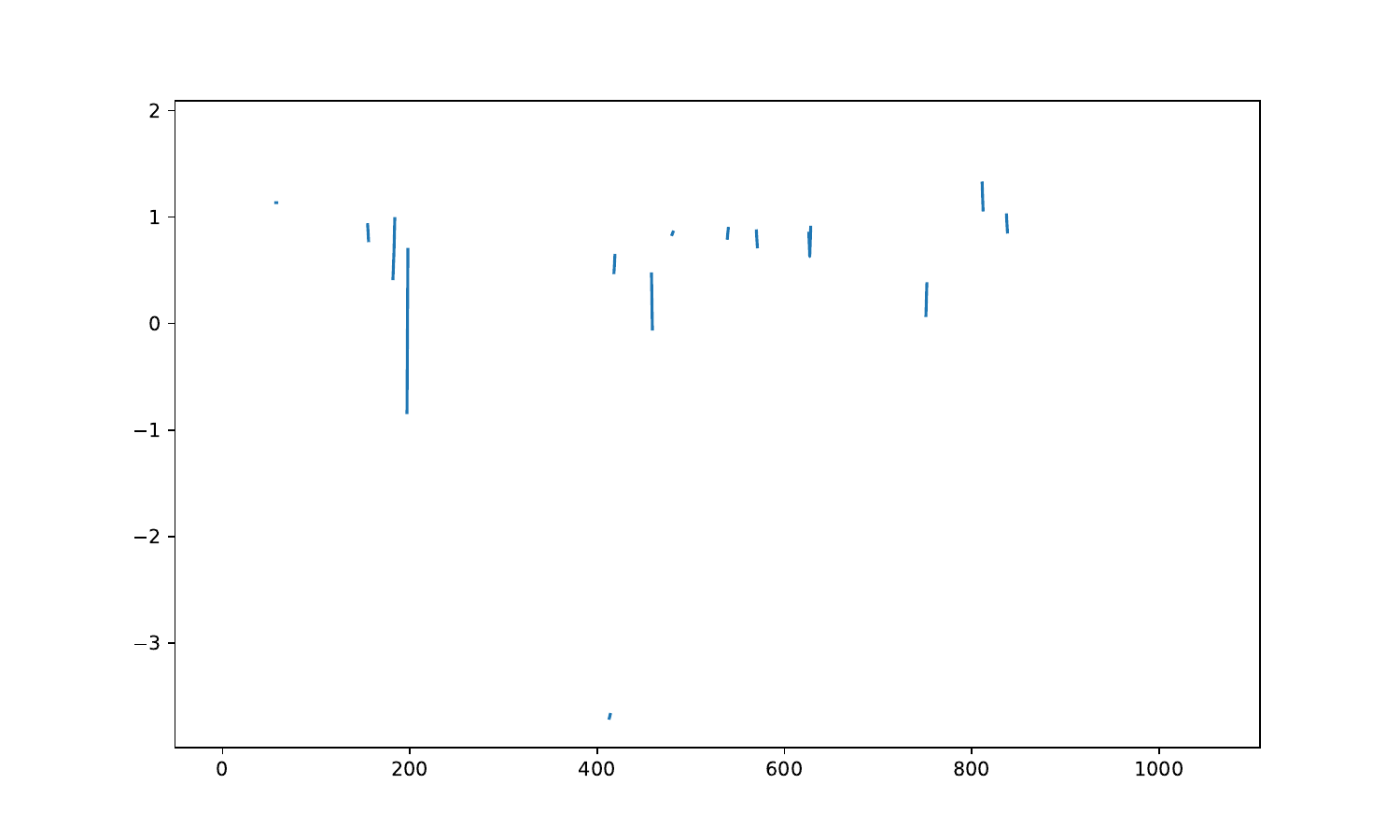}\label{fig: imputation90}}
    \subfigure[Imputation Result]{\includegraphics[width=0.45\textwidth]{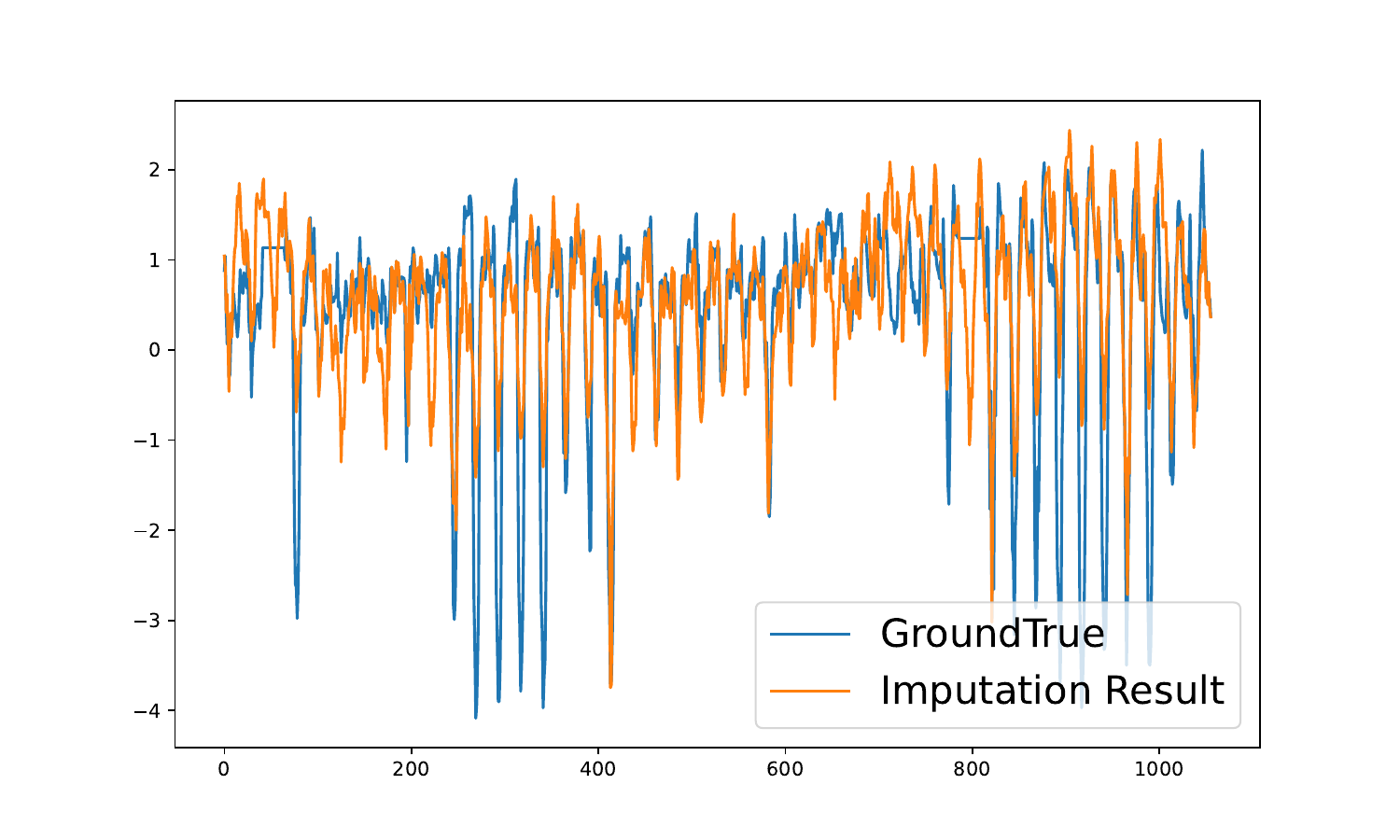}\label{fig: imputation_res90}}

    \caption{\textbf{Visualization of Imputation Results.}}
    \label{fig: imputation}
\end{figure*}

Time series imputation under the GPD paradigm is similar to time series forecasting. We also rely on a pre-trained model without fine-tuning. Given a sequence $Y$ with missing data, where the part of missing values can be considered as a mask $M$. We can use the non-missing data as prompts to align the distribution according to the method introduced in Section \ref{sec:prompt foecasting}. The imputation task can be completed simply by embedding the prompts according to the following formula:
\begin{equation}
    X_{t-1} = \{x^i_{t-1}\}^L_{i=1} * (1-M) + \{y^i_{t-1}\}^L_{i=1} * M
\end{equation}
where, $X_{t-1}$ is the posterior distribution of the $t$-th step in the reverse diffusion process, and $L$ represents the length of the entire sequence. We illustrate the imputation visualization with ratios of {30\%, 60\%, 90\%} in Fig. \ref{fig: imputation}. Our method demonstrates excellent performance even with 90\% of the values missing. Note that for additional downstream tasks, we do not introduce additional learning. Instead, we rely on the strong generalization and zero-shot capabilities of the GPD paradigm. Compared to end-to-end modeling methods, generative pre-training paradigms show more potential to become foundation models for time series.

\section{Zero-Shot Classification}
\label{classfication}

\citet{classifier} propose that diffusion models can serve as zero-shot classifiers. Following their method, we validate the zero-shot capability of GPD in classification task. Note that \citet{classifier} build upon multi-modal diffusion models, utilizing specific category text to modulate diffusion errors. Since GPD is not based on multimodality, we propose some enhancements and conduct simple binary classification experiments. We consider models trained on a single domain as experts in a Mixture of Experts (MoE \citep{masoudnia2014mixture,jacobs1991adaptive}), each proficient in specific domain knowledge. Specifically, pre-trained models based on long-range forecasting on the ETTh1 and Weather datasets perform classification experiments. Given a test sequence \( Y_0 \in \{ \text{ETTh1}, \text{Weather} \} \), we perform forward diffusion with reparameterization to add noise:

\begin{equation}
Y_t = \sqrt{\bar \alpha_t} Y_0 + \sqrt{1-\bar \alpha_t}\epsilon
\end{equation}

Then, we utilize these two pre-trained models to obtain predicted noise (\( \epsilon^E_\theta \) and \( \epsilon^W_\theta \)) at each time step. Subsequently, following Li et al.'s error computation method, we obtain the diffusion errors $E_E$=\( \text{min}(Errors_E) \) and $E_W$=\( \text{min}(Errors_W) \). The classification result for a specific domain is determined by the smaller domain error \text{min}([$E_E,E_W$]). We randomly sample 1000 time series from the testset of ETTh1 and Weather for evaluation. The classification accuracy on both datasets exceeds 90\%.

\section{More Experimental Details}

All our experiments are conducted in a completely fair manner. The only difference lies in the batch size settings. Since our model is smaller, we can use larger batch sizes for testing. The historical length settings in different works \citep{unitime,patchtst,timellm} result in different lengths of the testset. So we calculate the length of their testsets beforehand and adjust our batch size to ensure consistency in testset length. For instance, on the ETTh1 dataset with a historical length of 512 and a prediction length of 712, their testset length is 2161, thereby we set our batch size to 2161.

\section{Visualization}

\begin{figure*}[t]
    \centering
    \subfigure[w$/$o IN]{\includegraphics[width=0.48\textwidth]{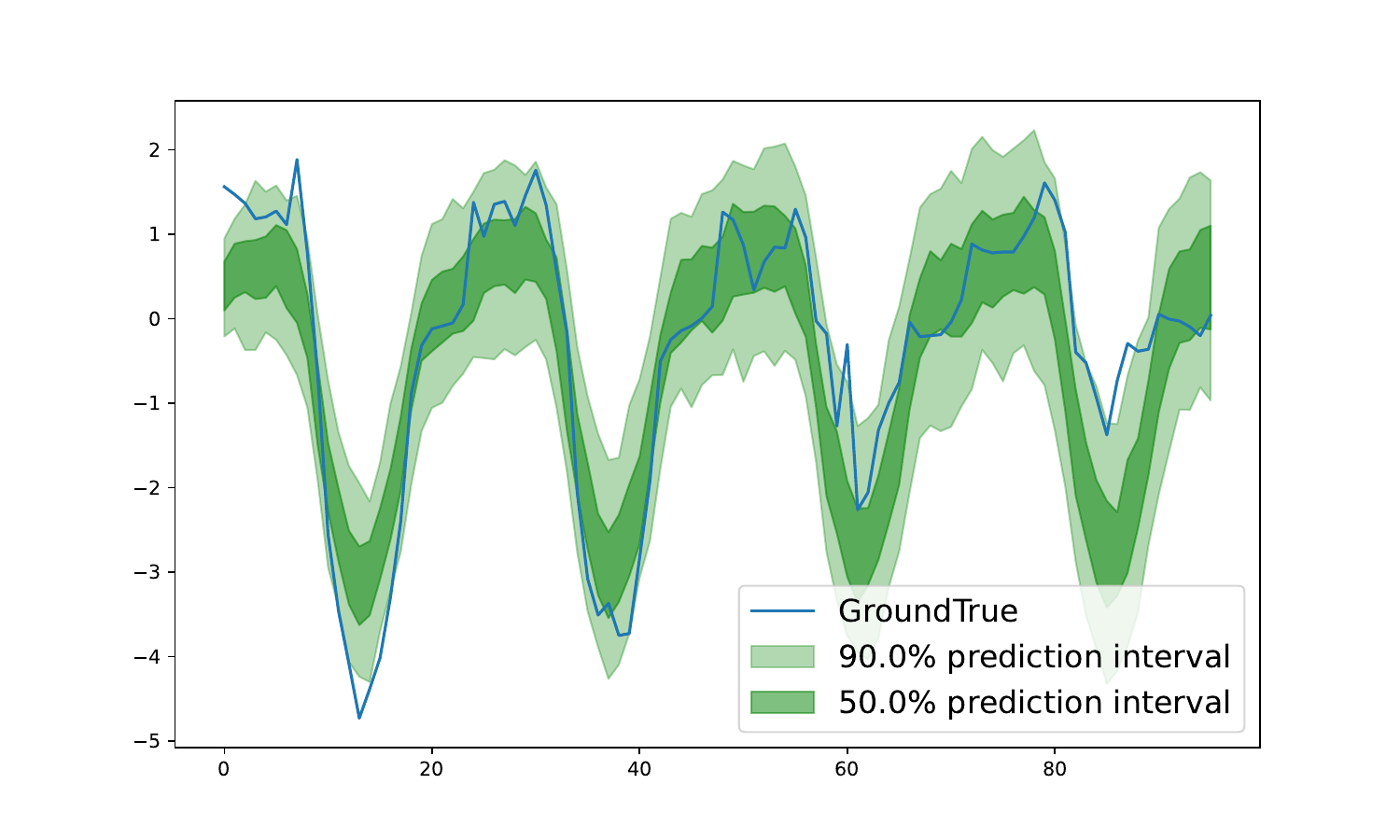}\label{fig: vis1-96-woIN}}
    \subfigure[SIN]{\includegraphics[width=0.48\textwidth]{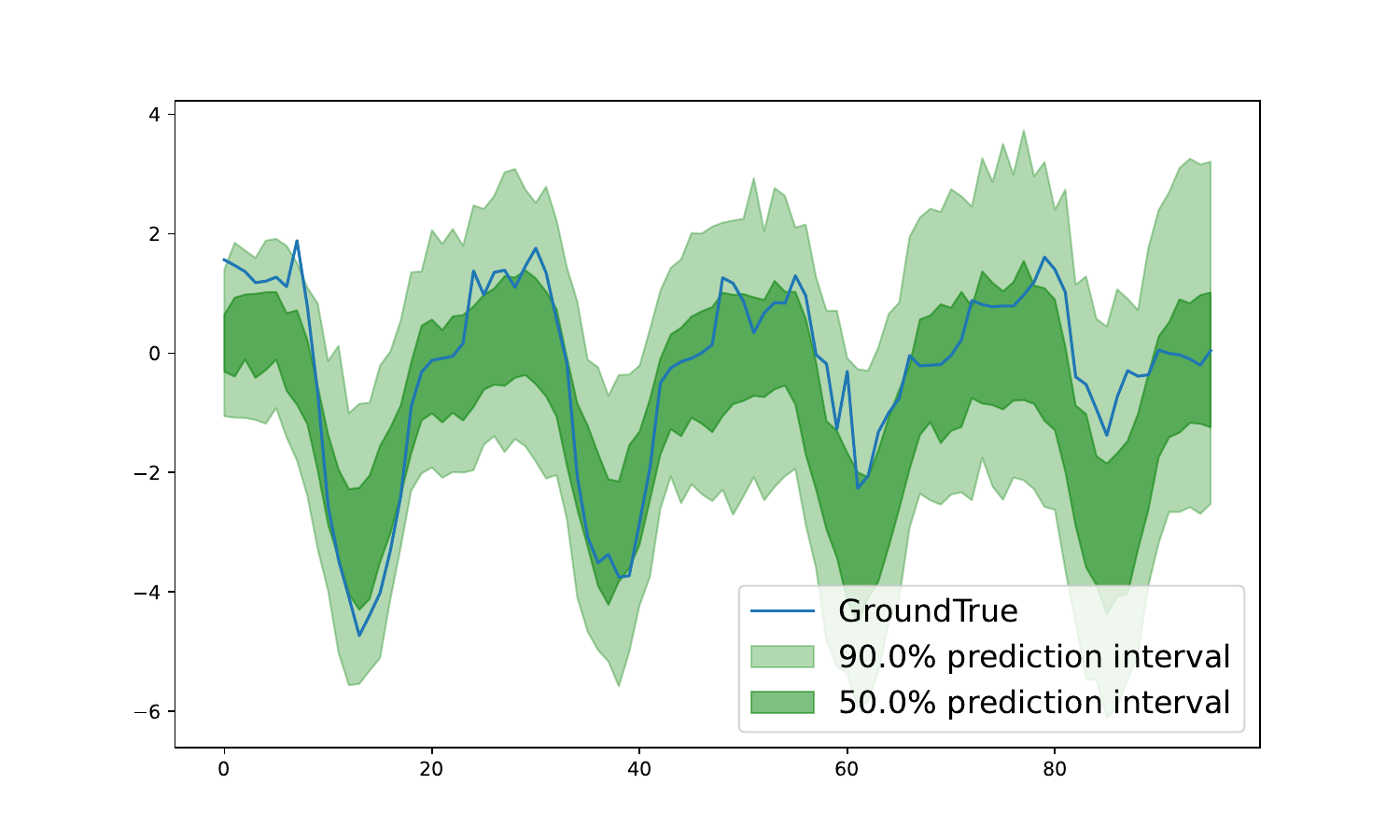}\label{fig: vis1-96-SIN}}
    
    \subfigure[w$/$o IN]{\includegraphics[width=0.48\textwidth]{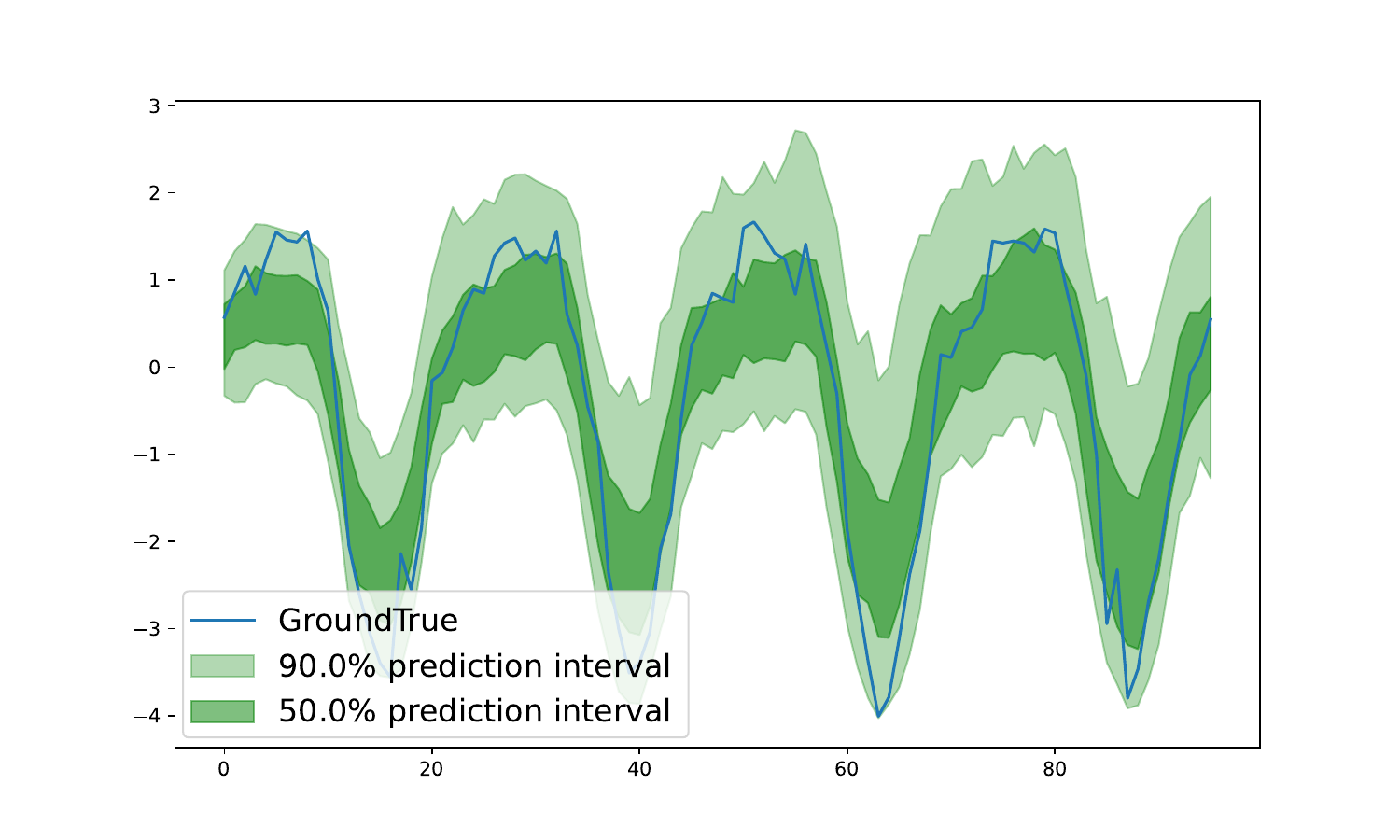}\label{fig: vis2-96-woIN}}
    \subfigure[SIN]{\includegraphics[width=0.48\textwidth]{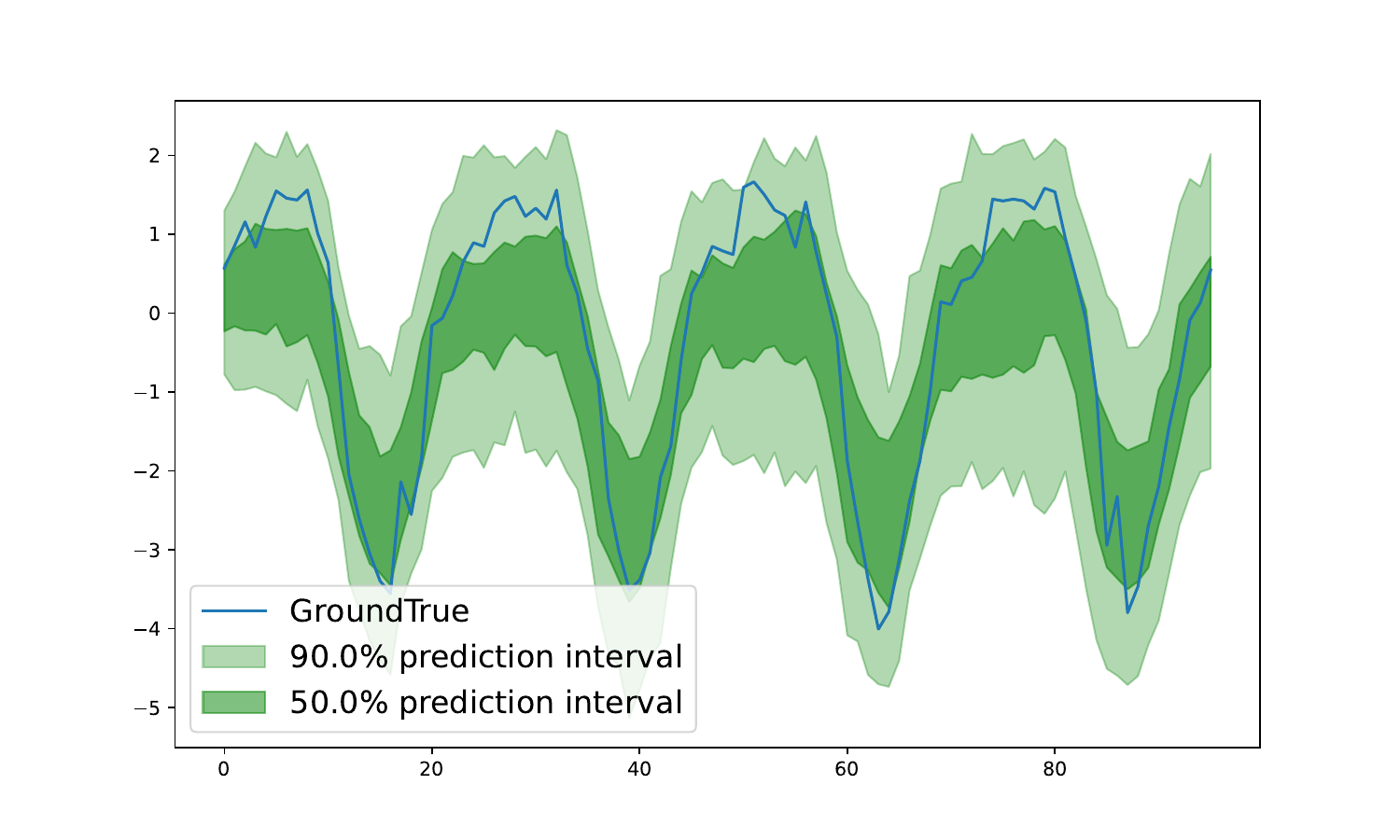}\label{fig: vis2-96-SIN}}
    \caption{Visualization comparison of predicted distributions for sampling instance normalization (SIN) and without instance normalization (w/o IN).}
    \label{fig:IN}
\end{figure*}

The forecasting results of pure generative models exhibit non-uniqueness. Therefore, we display the median, 50\%, and 90\% distribution intervals of 100 samples in Fig. \ref{fig:IN}. Simultaneously, we present visual comparisons of sampling instance normalization (SIN) against results without normalization. As shown in Fig. \ref{fig:IN}, the uncertainty of SIN is greater, compared to when no normalization is applied, hence it tends to predict more boldly in the face of abrupt trends.

\begin{table}[t]
\begin{center}
\caption{\textbf{Full long-term forecasting results}. A lower value indicates better performance. \textcolor{red}{Red} and \textcolor{blue}{blue} indicates the best and the second best performance.}
\label{tab:full-long-term-forecasting}
\begin{small}
\scalebox{0.7}{
\setlength \tabcolsep{3pt}
\begin{tabular}{c|c|cc|cc|cc|cc|cc|cc|cc|cc|cc|cc}
\toprule
\multicolumn{2}{c|}{Methods}&\multicolumn{2}{c|}{Ours}&\multicolumn{2}{c|}{TimeLLM}&\multicolumn{2}{c|}{LLM4TS}&\multicolumn{2}{c|}{GPT4TS}&\multicolumn{2}{c|}{DLinear}&\multicolumn{2}{c|}{PatchTST}&\multicolumn{2}{c|}{FEDformer}&\multicolumn{2}{c|}{Autoformer}&\multicolumn{2}{c|}{Informer}&\multicolumn{2}{c}{Pyraformer}\\
\midrule
\multicolumn{2}{c|}{Metric} & MSE & MAE & MSE & MAE & MSE & MAE & MSE & MAE & MSE & MAE & MSE & MAE & MSE & MAE& MSE & MAE & MSE & MAE & MSE & MAE \\
\midrule
\multirow{5}{*}{\rotatebox{90}{ETTh1}} & 
96  & \textcolor{red}{0.350} & \textcolor{red}{0.390} & \textcolor{blue}{0.362} & \textcolor{blue}{0.392} & 0.371 & 0.394 & 0.376 & 0.397 & 0.375 & 0.399 & 0.370 & 0.399 & 0.376 & 0.419 & 0.449 & 0.459 & 0.865 & 0.713 & 0.664 & 0.612 \\
& 192 & \textcolor{red}{0.375} & \textcolor{red}{0.409} & \textcolor{blue}{0.398} & 0.418 & 0.403 & \textcolor{blue}{0.412} & 0.416 & 0.418 & 0.405 & 0.416 & 0.413 & 0.421 & 0.420 & 0.448 & 0.500 & 0.482 & 1.008 & 0.792 & 0.790 & 0.681 \\
& 336 & \textcolor{red}{0.390} & \textcolor{blue}{0.424} & 0.430 & 0.427 & \textcolor{blue}{0.420} & \textcolor{red}{0.422} & 0.442 & 0.433 & 0.439 & 0.443 & 0.422 & 0.436 & 0.459 & 0.465 & 0.521 & 0.496 & 1.107 & 0.809 & 0.891 & 0.738 \\
& 720 & 0.450 & 0.467 & \textcolor{blue}{0.442} & \textcolor{blue}{0.457} & \textcolor{red}{0.422} & \textcolor{red}{0.444} & 0.477 & 0.456 & 0.472 & 0.492 & 0.447 & 0.466 & 0.506 & 0.507 & 0.514 & 0.512 & 1.181 & 0.865 & 0.963 & 0.782 \\
&Avg & \textcolor{red}{0.391} & \textcolor{blue}{0.423} & 0.408 & 0.423 & \textcolor{blue}{0.404} & \textcolor{red}{0.418} & 0.428 & 0.426 & 0.423 & 0.438 & 0.413 & 0.461 & 0.440 & 0.677 & 0.496 & 0.487 & 1.040 & 0.795 & 0.827 & 0.703\\
\midrule
\multirow{5}{*}{\rotatebox{90}{ETTh2}}
& 96  & \textcolor{red}{0.237} & \textcolor{red}{0.316} & 0.268 & \textcolor{blue}{0.328} & \textcolor{blue}{0.262} & 0.332 & 0.285 & 0.342 & 0.289 & 0.353 & 0.274 & 0.336 & 0.358 & 0.397 & 0.346 & 0.388 & 3.755 & 1.525 & 0.645 & 0.597 \\
& 192 & \textcolor{red}{0.272} & \textcolor{red}{0.344} & 0.329 & 0.375 & \textcolor{blue}{0.328} & \textcolor{blue}{0.377} & 0.354 & 0.389 & 0.383 & 0.418 & 0.339 & 0.379 & 0.429 & 0.439 & 0.456 & 0.452 & 5.602 & 1.931 & 0.799 & 0.683 \\
& 336 & \textcolor{red}{0.297} & \textcolor{red}{0.366} & 0.368 & 0.409 & 0.353 & 0.396 & 0.373 & 0.407 & 0.448 & 0.465 & \textcolor{blue}{0.329} & \textcolor{blue}{0.380} & 0.496 & 0.487 & 0.482 & 0.486 & 4.721 & 1.835 & 0.907 & 0.747 \\
& 720 & 0.394 & 0.434 & \textcolor{red}{0.372} & \textcolor{red}{0.420} & 0.383 & 0.425 & 0.406 & 0.441 & 0.605 & 0.551 & \textcolor{blue}{0.379} & \textcolor{blue}{0.422} & 0.463 & 0.474 & 0.515 & 0.511 & 3.647 & 1.625 & 0.963 & 0.783 \\
&Avg & \textcolor{red}{0.300} & \textcolor{red}{0.365} & 0.334 & 0.383 & \textcolor{blue}{0.331} & \textcolor{blue}{0.383} & 0.355 & 0.395 & 0.431 & 0.447 & 0.370 & 0.379 & 0.437 & 0.449 & 0.450 & 0.459 & 4.431 & 1.729 & 0.829 & 0.703 \\
\midrule
\multirow{5}{*}{\rotatebox{90}{ETTm1}}
& 96  & 0.288 & \textcolor{blue}{0.340} & \textcolor{red}{0.272} & \textcolor{red}{0.334} & \textcolor{blue}{0.285} & 0.343 & 0.292 & 0.346 & 0.299 & 0.343 & 0.290 & 0.342 & 0.379 & 0.419 & 0.505 & 0.475 & 0.672 & 0.571 & 0.543 & 0.510 \\
& 192 & 0.325 & \textcolor{blue}{0.363} & \textcolor{red}{0.310} & \textcolor{red}{0.358} & \textcolor{blue}{0.324} & 0.366 & 0.332 & 0.372 & 0.335 & 0.365 & 0.332 & 0.369 & 0.426 & 0.441 & 0.553 & 0.496 & 0.795 & 0.669 & 0.557 & 0.537\\
& 336 & 0.363 & 0.386 & \textcolor{red}{0.352} & \textcolor{red}{0.384} & \textcolor{blue}{0.353} & \textcolor{blue}{0.385} & 0.366 & 0.394 & 0.369 & 0.386 & 0.366 & 0.392 & 0.445 & 0.459 & 0.621 & 0.537 & 1.212 & 0.871 & 0.754 & 0.655 \\
& 720 & 0.421 & \textcolor{blue}{0.418} & \textcolor{red}{0.383} & \textcolor{red}{0.411} & \textcolor{blue}{0.408} & 0.419 & 0.417 & 0.421 & 0.425 & 0.421 & 0.416 & 0.420 & 0.543 & 0.490 & 0.671 & 0.561 & 1.166 & 0.823 & 0.908 & 0.724 \\
&Avg & 0.349 & \textcolor{blue}{0.377} & \textcolor{red}{0.329} & \textcolor{red}{0.372} & \textcolor{blue}{0.343} & 0.378 & 0.352 & 0.383 & 0.357 & 0.379 & 0.351 & 0.381 & 0.448 & 0.452 & 0.588 & 0.517 & 0.961 & 0.734 & 0.691 & 0.607 \\
\midrule
\multirow{5}{*}{\rotatebox{90}{ETTm2}}
& 96  & 0.168 & \textcolor{red}{0.253} & \textcolor{red}{0.161} & \textcolor{blue}{0.253} & \textcolor{blue}{0.165} & 0.254 & 0.173 & 0.262 & 0.167 & 0.269 & 0.165 & 0.255 & 0.203 & 0.287 & 0.255 & 0.339 & 0.365 & 0.453 & 0.435 & 0.507 \\
& 192 & 0.227 & \textcolor{red}{0.292} & \textcolor{red}{0.219} & 0.293 & \textcolor{blue}{0.220} & \textcolor{blue}{0.292} & 0.229 & 0.301 & 0.224 & 0.303 & 0.220 & 0.292 & 0.269 & 0.328 & 0.281 & 0.340 & 0.533 & 0.563 & 0.730 & 0.673\\
& 336 & 0.282 & \textcolor{blue}{0.327} & \textcolor{blue}{0.271} & 0.329 & \textcolor{red}{0.268} & \textcolor{red}{0.326} & 0.286 & 0.341 & 0.281 & 0.342 & 0.274 & 0.329 & 0.325 & 0.366 & 0.339 & 0.372 & 1.363 & 0.887 & 1.201 & 0.845 \\
& 720 & 0.372 & 0.385 & \textcolor{blue}{0.352} & \textcolor{red}{0.379} & \textcolor{red}{0.350} & \textcolor{blue}{0.380} & 0.378 & 0.401 & 0.397 & 0.421 & 0.362 & 0.385 &  0.421 & 0.415 & 0.433 & 0.432 & 3.379 & 1.338 & 3.625 & 1.451\\
&Avg & 0.262 & 0.314 & \textcolor{blue}{0.251} & \textcolor{blue}{0.313} & \textcolor{red}{0.251} & \textcolor{red}{0.313} & 0.267 & 0.326 & 0.267 & 0.334 & 0.255 & 0.315 & 0.305 & 0.349 & 0.327 & 0.371 & 1.410 & 0.810 & 1.498 & 0.869 \\
\midrule
\multirow{5}{*}{\rotatebox{90}{Weather}}
& 96  & 0.159 & 0.221 & \textcolor{blue}{0.147} & 0.201 & \textcolor{red}{0.147} & \textcolor{red}{0.196} & 0.162 & 0.212 & 0.176 & 0.237 & 0.149 & \textcolor{blue}{0.198} & 0.217 & 0.296 & 0.266 & 0.336 & 0.300 & 0.384 & 0.896 & 0.556 \\
& 192 & 0.204 & 0.267 & \textcolor{red}{0.189} & \textcolor{red}{0.234} & \textcolor{blue}{0.191} & \textcolor{blue}{0.238} & 0.204 & 0.248 & 0.220 & 0.282 & 0.194 & 0.241 & 0.276 & 0.336 & 0.307 & 0.367 & 0.598 & 0.434 & 0.622 & 0.624 \\
& 336 & 0.252 & 0.309 & 0.262 & \textcolor{blue}{0.279} & \textcolor{red}{0.241} & \textcolor{red}{0.277} & 0.254 & 0.286 & 0.265 & 0.319 & \textcolor{blue}{0.245} & 0.282 & 0.339 & 0.380 & 0.359 & 0.395 & 0.578 & 0.523 & 0.739 & 0.753\\
& 720 & 0.322 & 0.362 & \textcolor{red}{0.304} & \textcolor{red}{0.316} & \textcolor{blue}{0.313} & \textcolor{blue}{0.329} & 0.326 & 0.337 & 0.333 & 0.362 & 0.314 & 0.334 & 0.403 & 0.428 & 0.419 & 0.428 & 1.059 & 0.741 & 1.004 & 0.934 \\
&Avg & 0.234 & 0.289 & \textcolor{red}{0.225} & \textcolor{red}{0.257} & \textcolor{blue}{0.223} & \textcolor{blue}{0.260} & 0.237 & 0.271 & 0.249 & 0.300 & 0.226 & 0.264 & 0.310 & 0.360 & 0.338 & 0.382 & 0.634 & 0.521 & 0.815 & 0.717 \\
\midrule
\multirow{5}{*}{\rotatebox{90}{Electricity}}
& 96  & \textcolor{blue}{0.128} & 0.238 & 0.131 & 0.224 & \textcolor{red}{0.128} & \textcolor{blue}{0.223} & 0.139 & 0.238 & 0.140 & 0.237 & 0.139 & \textcolor{red}{0.222} & 0.193 & 0.308 & 0.201 & 0.317 & 0.274 & 0.368 & 0.386 & 0.449\\
& 192 & \textcolor{blue}{0.148} & 0.258 & 0.152 & 0.241 & \textcolor{red}{0.146} & \textcolor{red}{0.240} & 0.153 & 0.251 & 0.153 & 0.249 & 0.157 & \textcolor{blue}{0.240} & 0.201 & 0.315 & 0.222 & 0.334 & 0.296 & 0.386 & 0.386 & 0.443\\
& 336 & 0.169 & 0.279 & \textcolor{red}{0.160} & \textcolor{red}{0.248} & \textcolor{blue}{0.163} & \textcolor{blue}{0.258} & 0.169 & 0.266 & 0.169 & 0.267 & 0.163 & 0.259 & 0.214 & 0.329 & 0.231 & 0.338 & 0.300 & 0.394 & 0.378 & 0.443\\
& 720 & 0.214 & 0.317 & \textcolor{red}{0.192} & 0.298 & 0.200 & \textcolor{blue}{0.292} & 0.206 & 0.297 & 0.203 & 0.301 & \textcolor{blue}{0.197} & \textcolor{red}{0.290} & 0.246 & 0.355 & 0.254 & 0.361 & 0.373 & 0.439 & 0.376 & 0.445 \\
&Avg & 0.165 & 0.273 & \textcolor{red}{0.158} & \textcolor{red}{0.252} & \textcolor{blue}{0.159} & \textcolor{blue}{0.253} & 0.167 & 0.263 & 0.166 & 0.264 & 0.164 & 0.253 & 0.214 & 0.327 & 0.227 & 0.338 & 0.311 & 0.397 & 0.382 & 0.445\\
\midrule
\multirow{5}{*}{\rotatebox{90}{Traffic}}
& 96  & \textcolor{red}{0.359} & 0.260 & 0.362 & \textcolor{red}{0.248} & 0.372 & 0.259 & 0.388 & 0.282 & 0.420 & 0.282 & \textcolor{blue}{0.360} & \textcolor{blue}{0.249} & 0.587 & 0.366 & 0.613 & 0.388 & 0.719 & 0.391 & 2.085 & 0.468\\
& 192 & \textcolor{red}{0.373} & 0.267 & \textcolor{blue}{0.374} & \textcolor{red}{0.247} & 0.391 & 0.265 & 0.407 & 0.290 & 0.424 & 0.287 & 0.379 & \textcolor{blue}{0.256} & 0.604 & 0.373 & 0.616 & 0.382 & 0.696 & 0.379 & 0.867 & 0.467\\
& 336 & 0.399 & 0.282 & \textcolor{red}{0.385} & \textcolor{blue}{0.271} & 0.405 & 0.275 & 0.412 & 0.294 & 0.436 & 0.296 & \textcolor{blue}{0.392} & \textcolor{red}{0.264} & 0.621 & 0.383 & 0.622 & 0.337 & 0.777 & 0.420 & 0.869 & 0.469\\
& 720 & 0.440 & 0.305 & \textcolor{red}{0.430} & \textcolor{blue}{0.288} & 0.437 & 0.292 & 0.450 & 0.312 & 0.466 & 0.315 & \textcolor{blue}{0.432} & \textcolor{red}{0.286} & 0.626 & 0.382 & 0.660 & 0.408 & 0.864 & 0.472 & 0.881 & 0.473\\
&Avg & 0.393 & 0.279 & \textcolor{red}{0.388} & \textcolor{red}{0.264} & 0.401 & 0.273 & 0.414 & 0.295 & 0.437 & 0.264 & \textcolor{blue}{0.391} & \textcolor{blue}{0.264} & 0.611 & 0.376 & 0.628 & 0.379 & 0.764 & 0.416 & 1.176 & 0.469\\
\bottomrule
\end{tabular}
}
\end{small}
\end{center}
\end{table}

\begin{table}[t]
\begin{center}
\caption{\textbf{Full zero-shot learning results on ETT datasets}. A lower value indicates better performance. \textcolor{red}{Red} and \textcolor{blue}{blue} indicates the best and the second best performance.}
\label{tab:full-zero-shot-forecasting}
\begin{small}
\scalebox{0.75}{
\setlength \tabcolsep{3pt}
\begin{tabular}{c|c|cc|cc|cc|cc|cc|cc|cc|cc}
\toprule
\multicolumn{2}{c|}{Methods}&\multicolumn{2}{c|}{Ours}&\multicolumn{2}{c|}{TimeLLM}&\multicolumn{2}{c|}{LLMTime}&\multicolumn{2}{c|}{GPT4TS}&\multicolumn{2}{c|}{DLinear}&\multicolumn{2}{c|}{PatchTST}&\multicolumn{2}{c|}{TimesNet}&\multicolumn{2}{c}{Autoformer}\\
\midrule
\multicolumn{2}{c|}{Metric} & MSE & MAE & MSE & MAE & MSE & MAE & MSE & MAE & MSE & MAE & MSE & MAE& MSE & MAE & MSE & MAE \\
\midrule
\multirow{5}{*}{\rotatebox{0}{ETTh1} $\rightarrow$ \rotatebox{0}{ETTh2}} & 
96  & \textcolor{red}{0.239} & \textcolor{red}{0.321} & \textcolor{blue}{0.279} & \textcolor{blue}{0.337} &0.510 &0.576 & 0.335 & 0.374 & 0.347 & 0.400 & 0.304 & 0.350 & 0.358 & 0.387 & 0.469 & 0.486 \\
& 192 & \textcolor{red}{0.275} & \textcolor{red}{0.349} & \textcolor{blue}{0.351} & \textcolor{blue}{0.374} &0.523 &0.586 & 0.412 & 0.417 & 0.447 & 0.460 & 0.386 & 0.400 & 0.427 & 0.429 & 0.634 & 0.567 \\
& 336 & \textcolor{red}{0.301} & \textcolor{red}{0.371} & \textcolor{blue}{0.388} & \textcolor{blue}{0.415} &0.640 &0.637 & 0.441 & 0.444 & 0.515 & 0.505 & 0.414 & 0.428 & 0.449 & 0.451 & 0.655 & 0.588 \\
& 720 & \textcolor{blue}{0.396} & \textcolor{blue}{0.432} & \textcolor{red}{0.391} & \textcolor{red}{0.420} &2.296 &1.034 & 0.438 & 0.452 & 0.665 & 0.589 & 0.419 & 0.443 & 0.448 & 0.458 & 0.570 & 0.549 \\
&Avg & \textcolor{red}{0.303} & \textcolor{red}{0.368} & \textcolor{blue}{0.353} & \textcolor{blue}{0.387} &0.992 &0.708 & 0.406 & 0.422 & 0.493 & 0.488 & 0.380 & 0.405 & 0.421 & 0.431 & 0.582 & 0.548 \\
\midrule
\multirow{5}{*}{\rotatebox{0}{ETTh1} $\rightarrow$ \rotatebox{0}{ETTm2}}
& 96  & 0.224 & 0.316 & \textcolor{red}{0.189} & \textcolor{red}{0.293} &0.646&0.563& 0.236 & 0.315 & 0.255 & 0.357 & \textcolor{blue}{0.215} & \textcolor{blue}{0.304} & 0.239 & 0.313 & 0.352 & 0.432 \\
& 192 & 0.279 & 0.349 & \textcolor{red}{0.237} & \textcolor{red}{0.312} &0.934&0.654& 0.287 & 0.342 & 0.338 & 0.413 & \textcolor{blue}{0.275} & \textcolor{blue}{0.339} & 0.291 & 0.342 & 0.413 & 0.460 \\
& 336 & \textcolor{blue}{0.329} & 0.377 & \textcolor{red}{0.291} & \textcolor{red}{0.365} &1.157&0.728& 0.341 & 0.374 & 0.425 & 0.465 & 0.334 & 0.373 & 0.342 & \textcolor{blue}{0.371} & 0.465 & 0.489 \\
& 720 & \textcolor{blue}{0.413} & 0.426 & \textcolor{red}{0.372} & \textcolor{red}{0.390} & 4.730 &1.531 & 0.435 & 0.422 & 0.640 & 0.573 & 0.431 & 0.424 & 0.434 & \textcolor{blue}{0.419} & 0.599 & 0.551 \\
&Avg & \textcolor{blue}{0.311} & 0.367 & \textcolor{red}{0.273} & \textcolor{red}{0.340} & 1.867 &0.869 & 0.325 & 0.363 & 0.415 & 0.452 & 0.314 & \textcolor{blue}{0.360} & 0.327 & 0.361 & 0.457 & 0.483 \\
\midrule
\multirow{5}{*}{\rotatebox{0}{ETTh2} $\rightarrow$ \rotatebox{0}{ETTh1}}
& 96  & \textcolor{red}{0.394} & \textcolor{red}{0.421} & \textcolor{blue}{0.450} & \textcolor{blue}{0.452} &1.130 &0.777 & 0.732 & 0.577 & 0.689 & 0.555 & 0.485 & 0.465 & 0.848 & 0.601 & 0.693 & 0.569 \\
& 192 & \textcolor{red}{0.426} & \textcolor{red}{0.439} & \textcolor{blue}{0.465} & \textcolor{blue}{0.461} &1.242 &0.820 & 0.758 & 0.559 & 0.707 & 0.568 &0.565 & 0.509 & 0.860 & 0.610 & 0.760 & 0.601 \\
& 336 & \textcolor{red}{0.455} & \textcolor{red}{0.459} & \textcolor{blue}{0.501} & \textcolor{blue}{0.482} &1.328 &0.864 & 0.759 & 0.578 & 0.710 & 0.577 & 0.581 & 0.515 & 0.867 & 0.626 & 0.781 & 0.619 \\
& 720 & \textcolor{blue}{0.510} & \textcolor{blue}{0.505} & \textcolor{red}{0.501} & \textcolor{red}{0.502} &4.145 &1.461 & 0.781 & 0.597 & 0.704 & 0.596 & 0.628 & 0.561 & 0.887 & 0.648 & 0.796 & 0.644 \\
&Avg & \textcolor{red}{0.446} & \textcolor{red}{0.456} & \textcolor{blue}{0.479} & \textcolor{blue}{0.474} &1.961 & 0.981 & 0.757 & 0.578 & 0.703 & 0.574 & 0.565 & 0.513 & 0.865 & 0.621 & 0.757 & 0.608 \\
\midrule
\multirow{5}{*}{\rotatebox{0}{ETTh2} $\rightarrow$ \rotatebox{0}{ETTm2}}
& 96  & \textcolor{blue}{0.219} & \textcolor{blue}{0.306} & \textcolor{red}{0.174} & \textcolor{red}{0.276} &0.646&0.563& 0.253 & 0.329 & 0.240 & 0.336 & 0.226 & 0.309 & 0.248 & 0.324 & 0.263 & 0.352 \\
& 192 & \textcolor{blue}{0.280} & \textcolor{blue}{0.343} & \textcolor{red}{0.233} & \textcolor{red}{0.315} &0.934&0.654& 0.293 & 0.346 & 0.295 & 0.369 & 0.289 & 0.345 & 0.296 & 0.352 & 0.326 & 0.389 \\
& 336 & \textcolor{blue}{0.335} & \textcolor{blue}{0.376} & \textcolor{red}{0.291} & \textcolor{red}{0.337} &1.157&0.728& 0.347 & 0.376 & 0.345 & 0.397 & 0.348 & 0.379 & 0.353 & 0.383 & 0.387 & 0.426 \\
& 720 & \textcolor{blue}{0.429} & 0.430 & \textcolor{red}{0.392} & \textcolor{red}{0.417} &4.730 &1.531 & 0.446 & 0.429 & 0.432 & 0.442 & 0.439 & \textcolor{blue}{0.427} & 0.471 & 0.446 & 0.487 & 0.478 \\
&Avg & \textcolor{blue}{0.316} & \textcolor{blue}{0.364} & \textcolor{red}{0.272} & \textcolor{red}{0.341} &1.867 &0.869 & 0.335 & 0.370 & 0.328 & 0.386 & 0.325 & 0.365 & 0.342 & 0.376 & 0.366 & 0.411 \\
\midrule
\multirow{5}{*}{\rotatebox{0}{ETTm1} $\rightarrow$ \rotatebox{0}{ETTh2}}
& 96  & \textcolor{red}{0.280} & \textcolor{red}{0.348} & \textcolor{blue}{0.321} & \textcolor{blue}{0.369} &0.510 &0.576 & 0.353 & 0.392 & 0.365 & 0.415 & 0.354 & 0.385 & 0.377 & 0.407 & 0.435 & 0.470 \\
& 192 & \textcolor{red}{0.314} & \textcolor{red}{0.375} & \textcolor{blue}{0.389} & \textcolor{blue}{0.410} &0.523 &0.586 & 0.443 & 0.437 & 0.454 & 0.462 & 0.447 & 0.434 & 0.471 & 0.453 & 0.495 & 0.489 \\
& 336 & \textcolor{red}{0.332} & \textcolor{red}{0.393} & \textcolor{blue}{0.408} & \textcolor{blue}{0.433} &0.640 &0.637 & 0.469 & 0.461 & 0.496 & 0.494 & 0.481 & 0.463 & 0.472 & 0.484 & 0.470 & 0.472 \\
& 720 & \textcolor{red}{0.404} & \textcolor{blue}{0.439} & \textcolor{blue}{0.406} & \textcolor{red}{0.436} &2.296 &1.034 & 0.466 & 0.468 & 0.541 & 0.529 & 0.474 & 0.471 & 0.495 & 0.482 & 0.480 & 0.485 \\
&Avg & \textcolor{red}{0.333} & \textcolor{red}{0.389} & \textcolor{blue}{0.381} & \textcolor{blue}{0.412} &0.992 &0.708 & 0.433 & 0.439 & 0.464 & 0.475 & 0.439 & 0.438 & 0.457 & 0.454 & 0.470 & 0.479 \\
\midrule
\multirow{5}{*}{\rotatebox{0}{ETTm1} $\rightarrow$ \rotatebox{0}{ETTm2}}
& 96  & \textcolor{blue}{0.179} & \textcolor{blue}{0.260} & \textcolor{red}{0.169} & \textcolor{red}{0.257} &0.646&0.563& 0.217 & 0.294 & 0.221 & 0.314 & 0.195 & 0.271 & 0.222 & 0.295 & 0.385 & 0.457 \\
& 192 & \textcolor{blue}{0.231} & \textcolor{red}{0.293} & \textcolor{red}{0.227} & 0.318 &0.934&0.654& 0.277 & 0.327 & 0.286 & 0.359 & 0.258 & \textcolor{blue}{0.311} & 0.288 & 0.337 & 0.433 & 0.469 \\
& 336 & \textcolor{red}{0.288} & \textcolor{red}{0.332} & \textcolor{blue}{0.290} & \textcolor{blue}{0.338} &1.157&0.728& 0.331 & 0.360 & 0.357 & 0.406 & 0.317 & 0.348 & 0.341 & 0.367 & 0.476 & 0.477 \\
& 720 & \textcolor{red}{0.367} & \textcolor{blue}{0.381} & \textcolor{blue}{0.375} & \textcolor{red}{0.367} &4.730 &1.531 & 0.429 & 0.413 & 0.476 & 0.476 & 0.416 & 0.404 & 0.436 & 0.418 & 0.582 & 0.535 \\
&Avg & \textcolor{red}{0.266} & \textcolor{red}{0.317} & \textcolor{blue}{0.268} & \textcolor{blue}{0.320} &1.867 &0.869 & 0.313 & 0.348 & 0.335 & 0.389 & 0.296 & 0.334 & 0.322 & 0.354 & 0.469 & 0.484 \\
\midrule
\multirow{5}{*}{\rotatebox{0}{ETTm2} $\rightarrow$ \rotatebox{0}{ETTh2}}
& 96  & \textcolor{red}{0.253} & \textcolor{red}{0.325} & \textcolor{blue}{0.298} & \textcolor{blue}{0.356} &0.510 &0.576 & 0.360 & 0.401 & 0.333 & 0.391 & 0.327 & 0.367 & 0.360 &  0.401 & 0.353 & 0.393 \\
& 192 & \textcolor{red}{0.289} & \textcolor{red}{0.353} & \textcolor{blue}{0.359} & \textcolor{blue}{0.397} &0.523 &0.586 & 0.434 & 0.437 & 0.441 & 0.456 & 0.411 & 0.418 & 0.434 & 0.437 & 0.432 & 0.437 \\
& 336 & \textcolor{red}{0.313} & \textcolor{red}{0.375} & \textcolor{blue}{0.367} & \textcolor{blue}{0.412} &0.640 &0.637 & 0.460 & 0.459 & 0.505 & 0.503 & 0.439 & 0.447 & 0.460 & 0.459 & 0.452 & 0.459 \\
& 720 & \textcolor{red}{0.384} & \textcolor{red}{0.422} & \textcolor{blue}{0.393} & \textcolor{blue}{0.434} &2.296 &1.034 & 0.485 & 0.477 & 0.543 & 0.534 & 0.459 & 0.470 & 0.485 & 0.477 & 0.453 & 0.467 \\
&Avg & \textcolor{red}{0.311} & \textcolor{red}{0.369} & \textcolor{blue}{0.354} & \textcolor{blue}{0.400} &0.992 &0.708 & 0.435 & 0.443 & 0.455 & 0.471 & 0.409 & 0.425 & 0.435 & 0.443 & 0.423 & 0.439 \\
\midrule
\multirow{5}{*}{\rotatebox{0}{ETTm2} $\rightarrow$ \rotatebox{0}{ETTm1}}
& 96  & \textcolor{red}{0.345} & \textcolor{red}{0.380} & \textcolor{blue}{0.359} & \textcolor{blue}{0.397} &1.179& 0.781& 0.747 & 0.558 & 0.570 & 0.490 & 0.491 & 0.437 & 0.747 & 0.558 & 0.735 & 0.576 \\
& 192 & \textcolor{red}{0.364} & \textcolor{red}{0.393} & \textcolor{blue}{0.390} & \textcolor{blue}{0.420} &1.327&0.846& 0.781 & 0.560 & 0.590 & 0.506  & 0.530 & 0.470 & 0.781 & 0.560 & 0.753 & 0.586 \\
& 336 & \textcolor{red}{0.390} & \textcolor{red}{0.407} & \textcolor{blue}{0.421} & \textcolor{blue}{0.445} &1.478&0.902& 0.778 & 0.578 & 0.706 & 0.567  & 0.565 & 0.497 & 0.778 & 0.578 & 0.750 & 0.593 \\
& 720 & \textcolor{red}{0.437} & \textcolor{red}{0.434} & \textcolor{blue}{0.487} & \textcolor{blue}{0.488} &3.749 &1.408 & 0.769 & 0.573 & 0.731 & 0.584 & 0.686 & 0.565 & 0.769 & 0.573 & 0.782 & 0.609 \\
&Avg & \textcolor{red}{0.384} & \textcolor{red}{0.404} & \textcolor{blue}{0.414} & \textcolor{blue}{0.438} &1.933 &0.984 & 0.769 & 0.567 & 0.649 & 0.537 & 0.568 & 0.492 & 0.769 & 0.567 & 0.755 & 0.591 \\
\bottomrule
\end{tabular}
}
\end{small}
\end{center}
\end{table}

\begin{figure*}[t]
    \centering
    \subfigure[H=336, P=336]{\includegraphics[width=0.48\textwidth]{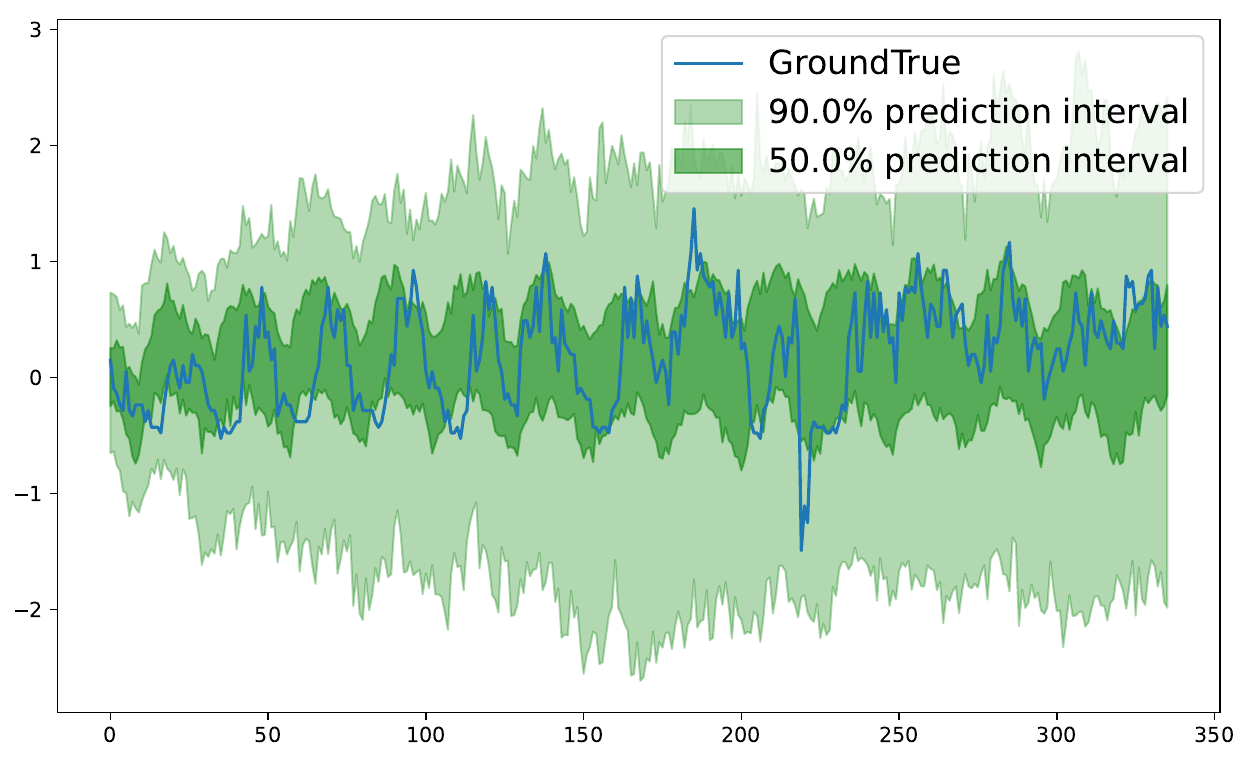}\label{fig: more-336-1}}
    \subfigure[H=336, P=336]{\includegraphics[width=0.48\textwidth]{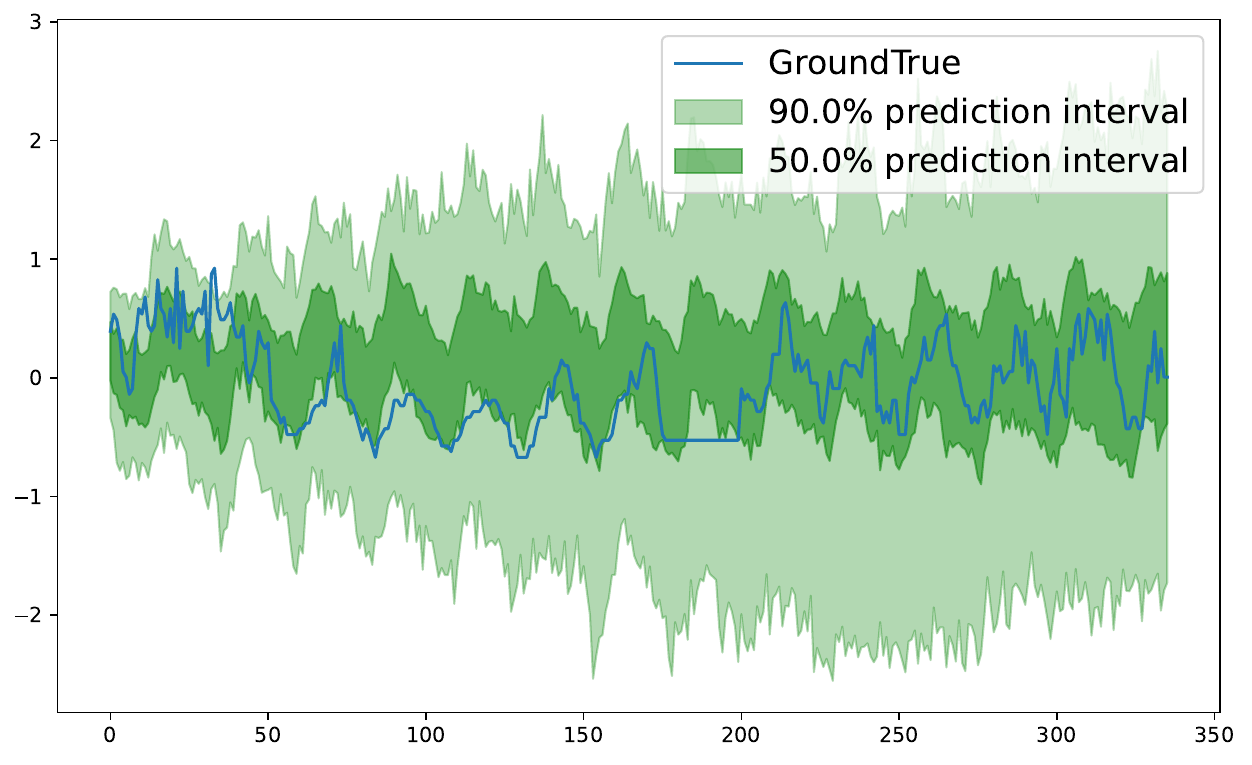}\label{fig: more-336-2}}

    \subfigure[H=336, P=336]{\includegraphics[width=0.48\textwidth]{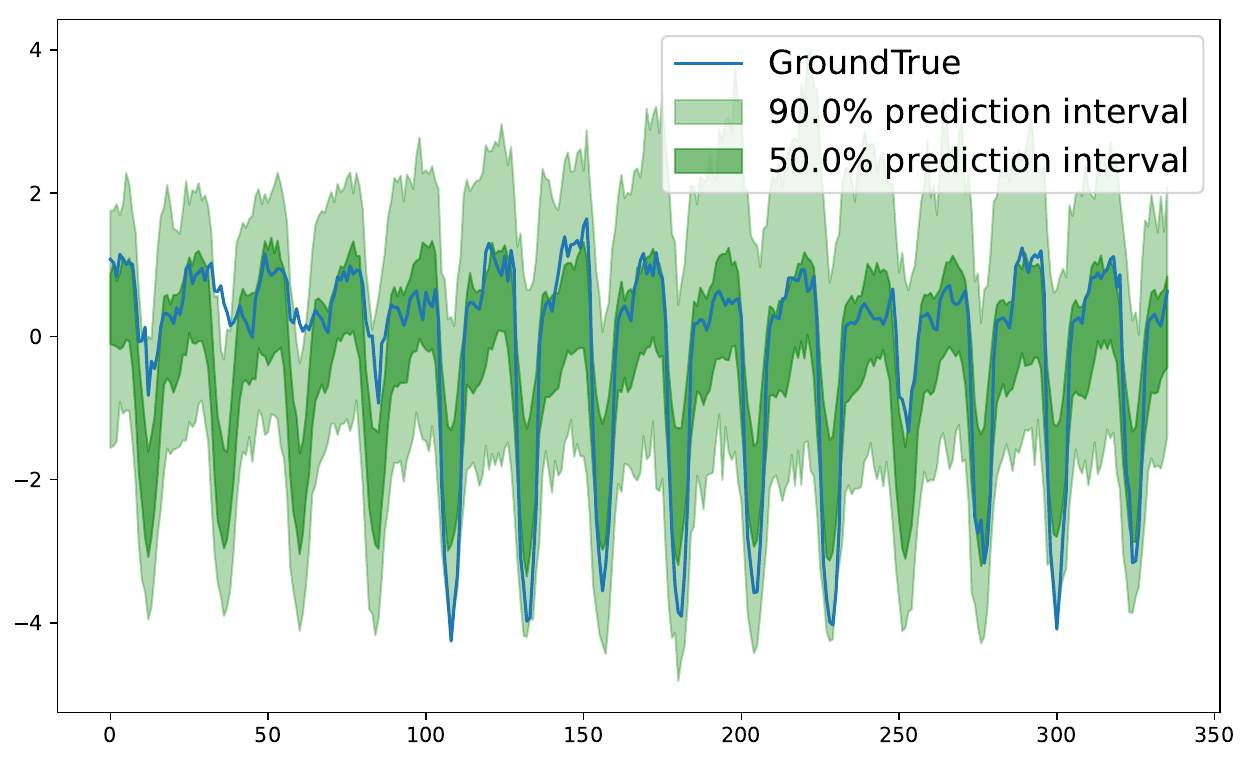}\label{fig: more-336-3}}
    \subfigure[H=336, P=336]{\includegraphics[width=0.48\textwidth]{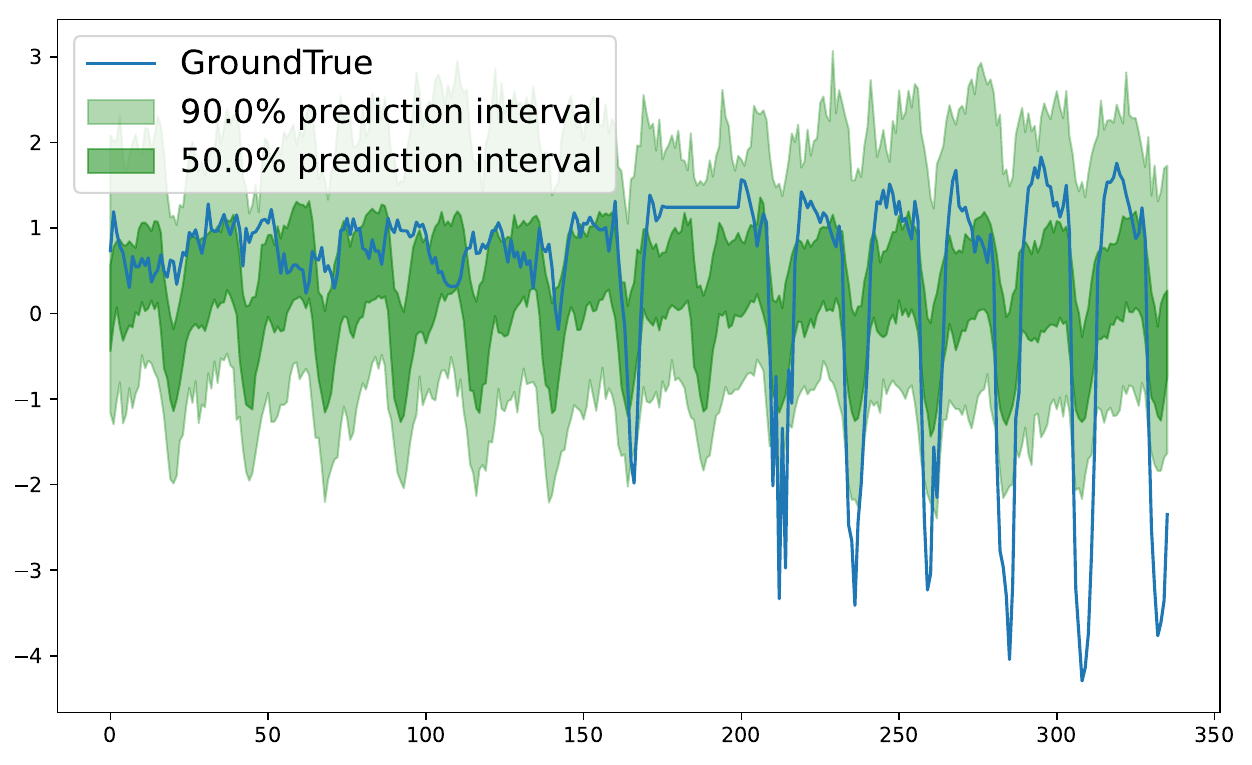}\label{fig: more-336-4}}

    \subfigure[H=336, P=720]{\includegraphics[width=0.48\textwidth]{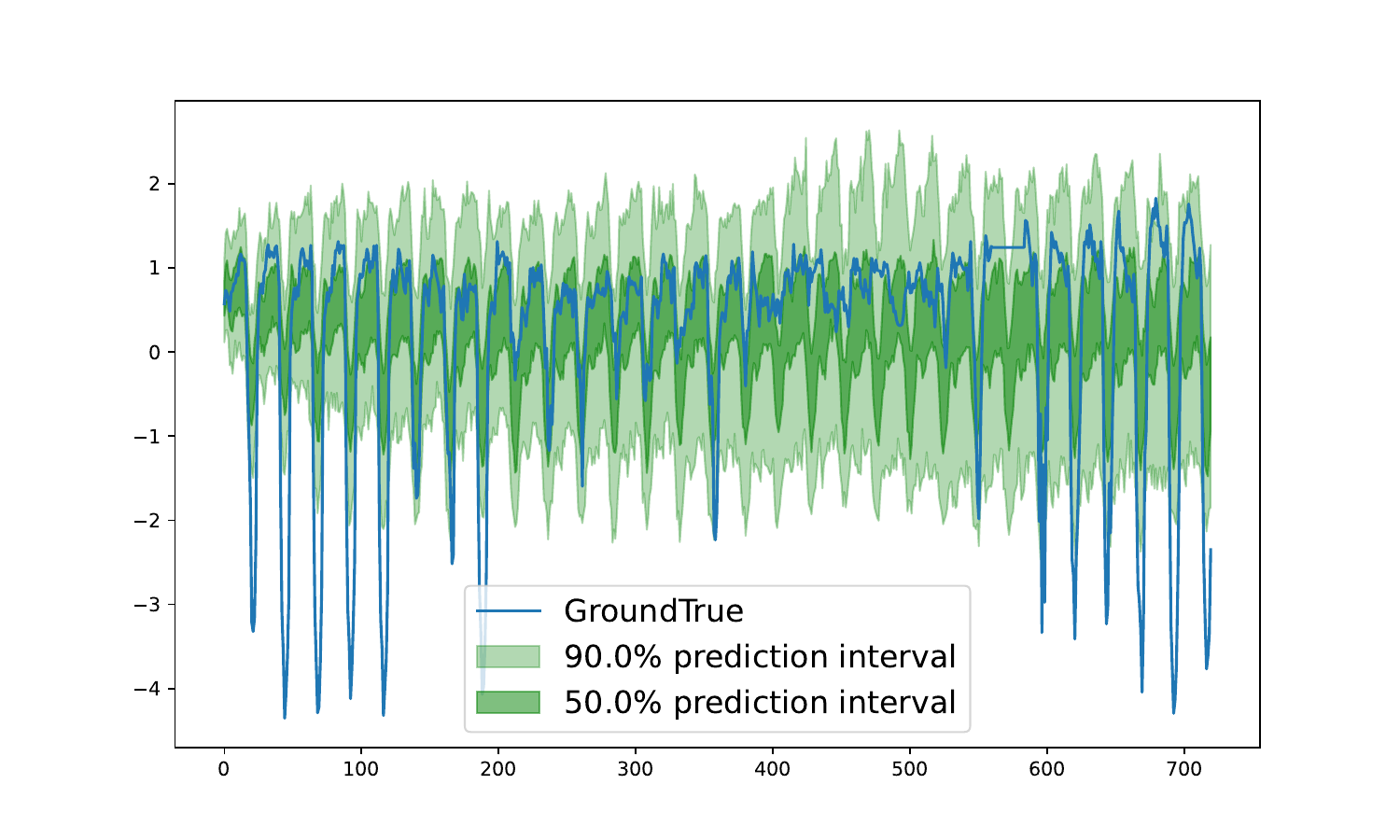}\label{fig: more-720-1}}
    \subfigure[H=336, P=720]{\includegraphics[width=0.48\textwidth]{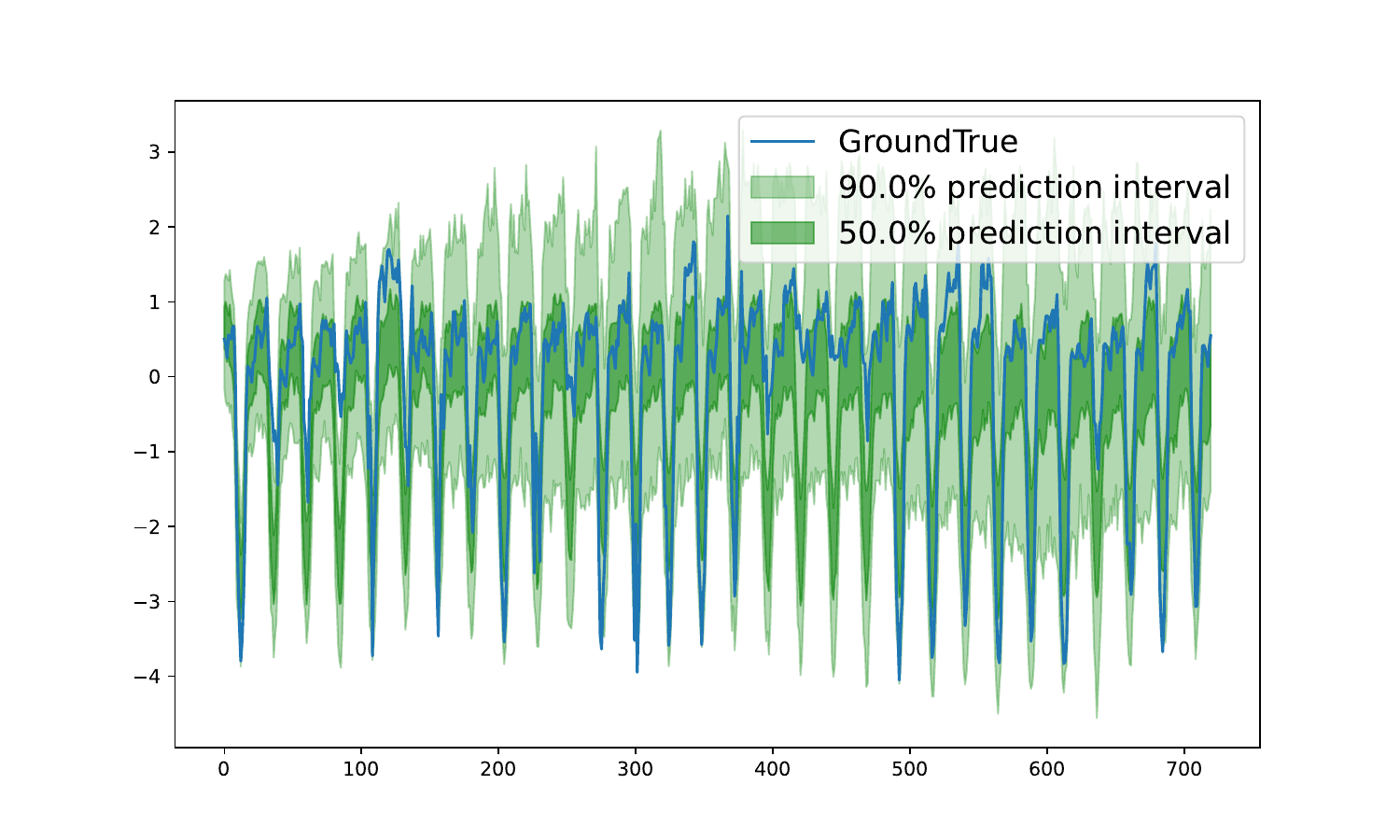}\label{fig: more-720-2}}
    
    \caption{More Results}
    \label{fig:more}
\end{figure*}

\newpage

\newpage

\end{document}